\documentclass[a4paper,11pt]{article}
\pdfoutput=1 

\usepackage{jheppub} 

\usepackage[T1]{fontenc} 

\usepackage{amsmath}                    
\usepackage[bbgreekl]{mathbbol}
\DeclareSymbolFontAlphabet{\mathbbl}{bbold}
\usepackage{amsfonts,amssymb,amsthm,mathtools}    
\DeclareSymbolFontAlphabet{\mathbbm}{bbold}
\DeclareSymbolFontAlphabet{\mathbb}{AMSb}%
\usepackage{url}
\usepackage{enumitem}
\usepackage{bm} 
\usepackage{tensor}
\usepackage[usenames,dvipsnames]{xcolor}
\usepackage{MnSymbol}
\usepackage{hyperref}
\usepackage{tensor}
\usepackage[titletoc,title]{appendix}
\usepackage{mathtools}
\usepackage{xcolor} 
\usepackage{indentfirst} 
\usepackage{scalerel}
\usepackage{graphicx}
\usepackage{todonotes}

\AtBeginDocument{}%

\makeatletter
\renewcommand*{\thesection}{\arabic{section}}

\newcommand\rrule[3][0pt]{%
	\ifdim#2>#3\math@hrule[#1]{#2}{#3}\else\math@vrule[#1]{#2}{#3}\fi}
\newcommand\math@hrule[3][0pt]{%
	\gdef\mystery@factor{0.07}%
	\@tempdima=#3%
	\rule[#1]{0pt}{#3}
	\raisebox{.5\@tempdima+#1}{%
		\makebox[#2][l]{\kern-.5\@tempdima\@@mathrule{#2}{#3}}}%
}
\newcommand\math@vrule[3][0pt]{%
	\gdef\mystery@factor{0.0}%
	\@tempdima=#2%
	\rule[#1]{0pt}{#3}
	\raisebox{-.0\@tempdima+#1}{%
		\kern0.5\@tempdima%
		\rotatebox{90}{\kern-0.5\@tempdima\makebox[#3][l]{\@@mathrule{#3}{#2}}}%
		\kern0.5\@tempdima}%
}
\def\@@mathrule#1#2{%
	\@tempdimb=#2%
	\@tempdima=\dimexpr#1-\mystery@factor\@tempdimb
	\pdfliteral{%
		q []0 d %
		1 J 
		\strip@pt\@tempdimb\space w \strip@pt\@tempdimb\space 0 m %
		\strip@pt\@tempdima\space 0 l S Q }}
\makeatother

\makeatletter
\DeclareFontFamily{OMX}{MnSymbolE}{}
\DeclareSymbolFont{MnLargeSymbols}{OMX}{MnSymbolE}{m}{n}
\SetSymbolFont{MnLargeSymbols}{bold}{OMX}{MnSymbolE}{b}{n}
\DeclareFontShape{OMX}{MnSymbolE}{m}{n}{
<-6>  MnSymbolE5
<6-7>  MnSymbolE6
<7-8>  MnSymbolE7
<8-9>  MnSymbolE8
<9-10> MnSymbolE9
<10-12> MnSymbolE10
<12->   MnSymbolE12
}{}
\DeclareFontShape{OMX}{MnSymbolE}{b}{n}{
<-6>  MnSymbolE-Bold5
<6-7>  MnSymbolE-Bold6
<7-8>  MnSymbolE-Bold7
<8-9>  MnSymbolE-Bold8
<9-10> MnSymbolE-Bold9
<10-12> MnSymbolE-Bold10
<12->   MnSymbolE-Bold12
}{}

\let\llangle\@undefined
\let\rrangle\@undefined
\DeclareMathDelimiter{\llangle}{\mathopen}%
{MnLargeSymbols}{'164}{MnLargeSymbols}{'164}
\DeclareMathDelimiter{\rrangle}{\mathclose}%
{MnLargeSymbols}{'171}{MnLargeSymbols}{'171}
\makeatother

\def\wwedgee{{\setbox0\hbox{\ensuremath{\mathrel{\wedge}}}\rlap{\hbox to \wd0{\hss\hspace*{.6ex}\ensuremath\wedge\hss}}\box0}}

\newcommand{\www}{\mathrel{\wwedgee}}

\newcommand{\bd}{\bm{\mathrm{d}}}
\newcommand{\bD}{\bm{\mathrm{D}}}

\def\lcdr{\left(\cdot\right)}

\newcommand{\be}{\bm{\mathrm{e}}}
\newcommand{\bh}{\bm{\mathrm{h}}}
\newcommand{\bomega}{\bm{\mathrm{\omega}}}
\newcommand{\bB}{\bm{\mathrm{B}}}
\newcommand{\bF}{\bm{\mathrm{F}}}
\newcommand{\bG}{\bm{\mathrm{G}}}
\newcommand{\balpha}{\bm{\mathrm{\alpha}}}

\newcommand{\dd}{\mathchoice
	{\mathbbm{d}\rrule{.087ex}{1.605ex}\hspace*{0.15ex}} 
	{\mathbbm{d}\rrule{.087ex}{1.605ex}\hspace*{0.15ex}} 
	{\mathbbm{d}\rrule{.08ex}{1.125ex}\hspace*{0.15ex}}  
	{\mathbbm{d}\rrule{.06ex}{.8ex}\hspace*{0.15ex}}     
}

\newtheorem{theorem}{Theorem}[section]
\newtheorem{lemma}[theorem]{Lemma}

\title{\boldmath A title with some math: $x=1$}

\author[a,b]{J. Fernando Barbero G.,}
\author[c]{Marc Basquens,}
\author[b, c]{Bogar Díaz,}
\author[b, c]{and Eduardo J. S. Villase\~nor.}

\affiliation[a]{Instituto de Estructura de la Materia, CSIC, Serrano 123, 28006 Madrid, Spain}
\affiliation[b]{Grupo de Teor\'{\i}as de Campos y F\'{\i}sica Estad\'{\i}stica, Instituto Gregorio Mill\'an (UC3M), Unidad Asociada al Instituto de Estructura de la Materia, CSIC}
\affiliation[c]{Departamento de Matem\'aticas, Universidad Carlos III de Madrid, Avda.\  de la Universidad 30, 28911 Legan\'es, Spain}

\emailAdd{fbarbero@iem.cfmac.csic.es}
\emailAdd{marc.basquens.munoz@gmail.com}
\emailAdd{bodiazj@math.uc3m.es}
\emailAdd{ejsanche@math.uc3m.es}

\abstract{We study the internally abelianized version of a range of gravitational theories, written in connection tetrad form, and study the possible interaction terms that can be added to them in a consistent way. We do this for 2+1 and 3+1 dimensional models. In the latter case we show that the Cartan-Palatini and Holst actions are not consistent deformations of their abelianized versions. We also show that the Husain-Kucha\v{r} and Euclidean self-dual actions are consistent deformations of their abelianized counterparts. This suggests that if the latter can be quantized, it could be possible to devise a perturbative scheme leading to the quantization of Euclidean general relativity along the lines put forward by Smolin in the  early nineties.}

\arxivnumber{2202.07997}

\begin{document}
	
\title{Consistent and non-consistent deformations of gravitational theories}

\maketitle
\flushbottom

\section{Introduction}

Soon after the introduction of the Ashtekar variables for General Relativity, Smolin outlined a quantum gravity program based on the complex self-dual action \cite{Smolin}. The main idea of that proposal was to consider Newton's constant $G$ as a coupling similar to the one appearing in Yang-Mills. By doing so the self-dual action could be split in two terms dependent and independent of $G$ respectively. This split is reminiscent of the standard one used in perturbative quantum gravity where, by writing the metric as the Minkowski metric plus a ``linear perturbation'', it is possible to expand the action of general relativity as a quadratic (``free'') part plus interaction terms proportional to powers of $G$. 

There are, however, a number of important differences between the two approaches. In the usual perturbative setting, a specific fixed metric is introduced. Despite being an auxiliary, background object put in by hand, its particular properties play a central role. For instance, if the Minkowski metric is used, its Poincaré symmetry allows us to interpret the free Lagrangian as describing massless spin-2 particles (gravitons) and the remaining terms as (non-renormalizable) interactions. 
In terms of the Minkowski metric and its perturbations, the gravitational Lagrangian has a very complicated expression consisting of a free part and an infinite number of interaction terms. Furthermore, none of these terms is diff-invariant by itself (although, in a definite sense, the full action is!). The standard perturbative approach ultimately fails due to the non-renormalizability of the theory.

On the other hand, in Smolin's proposal no auxiliary objects are introduced at any stage. Also, when the self-dual action is split, only two terms arise and both of them are diff-invariant. Smolin mentions in \cite{Smolin} that the integrability of the abelianized model could signal the possibility of quantizing it in an exact way and serve as the basis of a perturbative approach (although he does not provide a proof of this statement). In the present paper, we would like to take over Smolin's idea and study it in detail. We will pause and discuss a number of issues that are relevant in this context. 

In \cite{Smolin}, Smolin explicitly introduced the gravitational constant $G$ in the curvature and covariant exterior derivative: 
\begin{align*}
    \bF^{KL} &= \bd \tensor{\bomega}{^K^L} + G\, \tensor{\bomega}{^K_M} \wedge \tensor{\bomega}{^M^L} \ , \\
    \bD \be^I &= \bd \be^I + G\, \tensor{\bomega}{^I_J} \wedge \be^J \ ,
\end{align*}
which then yields the action
\begin{align*}
    S =  \int_{\mathcal{M}} \epsilon_{IJKL} \be^I \wedge \be^J \wedge \left( \bd \tensor{\bomega}{^K^L} + G\, \tensor{\bomega}{^K_M} \wedge \tensor{\bomega}{^M^L}\right) \ ,
\end{align*}
(here $\bomega$ is a self-dual connection 1-form) and then took $G=0$ in these expressions. Barros e Sá and Bengtsson considered the $G=0$ action for a generic $SO(1,3)$ connection \cite{BarrosESa}. More recently, these actions have been revisited by Bakhoda and Thiemann in \cite{Bakhoda_2021}.
A first relevant issue is that it is unclear what taking Newton's $G$ to $0$ means, let alone how it can be used as a perturbative parameter, since this is a \textit{dimensionful} quantity.  Moreover, this coupling constant is introduced by hand without providing a clear reason as to why this is the right way to proceed. 

An alternative proposal is to \textit{internally abelianize} the theory: i.e. to replace the internal symmetry group by an abelian one. This is actually what one does in the case of Yang-Mills: the free part consists of several copies of the Maxwell action and its group of local symmetries is just the direct product of those of the individual Maxwell terms. The abelianization has the sought effect on the model. This approach is sensible \textit{a priori} and useful \textit{a posterior}, since it yields nice and simple unperturbed models while being natural and without arbitrariness. 

A second problem with Smolin's scheme is that the integrability of the unperturbed model may not be enough to produce a working perturbative theory.  The standard perturbative treatment consisting in approximating the true solution by an asymptotic expansion in the perturbation parameter is only valid for \textit{regular perturbations}. Singular perturbative problems can still be tackled, of course,  but each of them may require its own particular approach.

The interaction terms that yield regular perturbations in field theories are called \textit{consistent deformations}, and the problem of finding all of them can be phrased in a comohological form \cite{Henneaux_Review, Henneaux}. Consistent deformations may change the gauge symmetries of the theory and their algebra, but they preserve their ``number''.
This is enforced by construction to ensure that the number of degrees of freedom does not change. Notice that, were this not true, a regular perturbation would be impossible, since solutions would appear and disappear. Hence, requiring that the interactions are consistent deformations of the internally abelianized theory is crucial for Smolin's program.

In a somewhat different direction, it is important to mention that a related approach has been followed in \cite{Tomlin, Varadarajan} to address the very difficult problem of quantizing the Hamiltonian constraint in Loop Quantum Gravity. The idea there is to first deal with the so-called $U(1)^3$ models, which are essentially those obtained by abelianizing the general relativity constraints in the Ashtekar formulation.

The paper is structured as follows. In Section \ref{section:strategy} we review the techniques that we are going to use in the paper.
Then, in Sections \ref{2+1} and \ref{3+1}, we explore a variety of models in $2+1$ and $3+1$ dimensions, study their internal abelianization and consistent deformations. We end the paper with our conclusions and several appendices where we provide some auxiliary results.

A few words on notation are in order. If $\mathsf{vol}$ is a volume form and $\bm\eta$ a top form, we denote by
\begin{align*}
    \displaystyle \left(\frac{\bm\eta}{\mathsf{vol}}\right)
\end{align*}
the function $\varphi$ satisfying ${\bm\eta}=\varphi\, \mathsf{vol}$.
We write the volume form defined by a nondegenerate frame $\be^I$ as
\begin{equation}\label{vole}
\mathsf{vol}_{\be}:=\frac{1}{4!}\epsilon_{IJKL}{\bm{\mathrm{e}}}^I\wedge {\bm{\mathrm{e}}}^J\wedge {\bm{\mathrm{e}}}^K\wedge {\bm{\mathrm{e}}}^L \ .
\end{equation}
We denote by $\mathcal{M}$ the spacetime, that we take without boundary and diffeomorphic to $\mathbb{R} \times \Sigma$.
We write the spacetime exterior derivative as $\bd$, the spacetime covariant exterior derivative as $\bD$, and the exterior derivative in the space of fields as $\dd$. The rest of the notation is standard.

\section{Strategy} \label{section:strategy}

In order to set up a perturbative scheme for a field theory ---both at the classical and quantum levels---, it must satisfy some conditions. First of all, one needs an appropriate unperturbed model. This particular choice is important as the viabibility of the scheme usually depends on its properties. In particular, it is natural to demand that the unperturbed theory be integrable in order to have explicit solutions that can be used as zeroth order terms of the perturbative expansions for solutions to the complete model.

Another necessary condition is that the chosen unperturbed model can be ``continuously'' deformed to the full theory, that is, that the perturbation is regular. This is so because perturbed theories that have a different number of degrees of freedom than their unperturbed counterparts are expected to behave in a singular way.
Intuitively, this happens because solutions are added or lost as the perturbative parameter changes, yielding a discontinuous process. In the context of gauge theories, such good perturbations are called \textit{consistent deformations}. They are required to preserve the number of gauge symmetries of the action but are allowed to change their form and their algebra. The standard technique to study them is the BRST formalism \cite{Henneaux_Teitelboim, Henneaux_Review}. Notice that, in this scheme, one assumes that the topology of $\mathcal{M}$ is trivial. Nonetheless, when we solve some of the equations of motion throughout the paper, we will be able to consider general topologies at almost no extra cost. Note that this is independent of the BRST method and we will do so for completeness.

Let us define the Euler forms $E_{\varphi A}$ by
\begin{align*}
    \dd S = \int_{\mathcal{M}} E_{\varphi A} \wedge \dd \varphi^A \ ,
\end{align*}
where the $\varphi^A$ are the dynamical fields, and $A$ denotes all possible internal indices. The gauge invariant directions correspond to the kernel of $\dd S$. Henceforth these gauge transformations will be denoted as $\mathbb{Z}_\epsilon$ and their components as $Z^A_\epsilon$, where $\epsilon$ stands for the gauge parameter.

When one deforms an action $S^{0}$ by introducing an interaction term $g S^{1}$
\begin{align*}
    S = S^{0} + g S^{1} \ ,
\end{align*}
the gauge symmetries of the resulting action ---as well as their algebra--- get deformed in the following way (we write $\mathbb{Z}_\epsilon = \mathbb{Z}_\epsilon^{0} + g \mathbb{Z}_\epsilon^{1}$): up to first order, we have
\begin{align}
\label{eq:deformed_symmetry}
    0 =&\,\, \dd{S}(\mathbb{Z}_\epsilon) = \int_{\mathcal{M}} E_{\varphi A} \wedge Z^A_\epsilon   \nonumber\\
    =&\int_{\mathcal{M}} \left( E_{\varphi A}^{0} \wedge \tensor{Z}{_\epsilon^{0}^A} +  g \left( E_{\varphi A}^{1} \wedge \tensor{Z}{_\epsilon^{0}^A} + E_{\varphi A}^{0} \wedge \tensor{Z}{_\epsilon^{1}^A}  \right) + O(g^2) \right)\ .
\end{align}
At zeroth order in $g$, \eqref{eq:deformed_symmetry} is automatically satisfied, however at first order this is not the case. We call the deformations for which the first order also vanishes \textit{consistent deformations}. Notice that, for a consistent deformation of an action $S^{0}$, $\dd{S}^{1}(\mathbb{Z}^{0}_\epsilon)$ must vanish when the unperturbed equations of motion $E_{\varphi A}^{0}$ hold (we will refer to this as `on-shell' and denote it as $\approx 0$).
Furthermore,
if the deformation is consistent, we can write
\begin{align}
\label{eq:deformation_eom}
   g\int_{\mathcal{M}} E_{\varphi A}^{1} \wedge \tensor{Z}{_\epsilon^{0}^A} = -g\int_{\mathcal{M}}  E_{\varphi A}^{0} \wedge \tensor{Z}{_\epsilon^{1}^A}\,.
\end{align}
It is usually easy to read off the deformation of the symmetries from this expression.
Notice that, if the deformation preserves the gauge symmetries, i.e. $\tensor{Z}{_\epsilon^{1}^A} = 0$, then $\dd{S}^{1}(\mathbb{Z}^0_\epsilon)$ vanishes identically. 
It is interesting to keep in mind that it is relevant which action we take as the starting point and which one we treat as a deformation, since the results may differ. We give an example of this in Appendix \ref{deformation_symmetry}.

The conditions discussed above can be translated into an appropriate cohomological language in such a way that it is possible to transform the problem of finding all the consistent deformations into that of computing the so-called BRST cohomology, which encodes all the information about observables. In general this is a hard problem, although it has been solved for some interesting models  \cite{Henneaux_Review, Henneaux, Boulanger2001, Boulanger2018, Boulanger2018b, Dai2020}. However, if one only wants to know whether a given interaction is a consistent deformation of an action, the problem reduces to just checking that \eqref{eq:deformed_symmetry} holds. 

In this paper we will use the following strategy based on some comments by Smolin in \cite{Smolin}. 
Given a field theory defined by an action, we will consider its internal abelianization: if the original group on which the connection is modelled is $\mathcal{G}$, its internal abelianization is obtained by using a $(\text{dim } \mathcal{G})$-dimensional abelian group instead of $\mathcal{G}$ as the starting point. 
This amounts to modifying the Lie algebra brackets of $\mathfrak{g}$ by a factor of $g$ (that could serve later on as a perturbative parameter) and letting $g \rightarrow 0$.
Then we check whether the original theory is a consistent deformation of the internally abelianized version. 
If this is the case---provided that the internally abelianized model is integrable---then a regular perturbative scheme is possible and could be pursued.

\section{Deformations in three spacetime dimensions}  \label{2+1}
Before addressing the physical case of four dimensions, it is worth devoting some attention to the three-dimensional case.
It has been known for some time that, by choosing an appropriate group, the Chern-Simons action generates many actions of interest in three dimensions. 
For instance, a $\mathcal{G}-$Cartan-Palatini-like theory is equivalent to a $I\mathcal{G}-$Chern-Simons model \cite{Romano, Achucarro-Townsend, Witten}:
\begin{align*}
    S_{\textrm{CS}}(A) = \frac{1}{2}\int_{\mathcal{M}} \langle A, \bd A + \frac{1}{3} [A, A] \rangle = \int_{\mathcal{M}}  \tensor{e}{_a} \wedge \tensor{F}{^a} = S_{\textrm{P}}(e, \omega) \ ,   
\end{align*}
where $A = e_a P^a + \omega^a J_a$, $F^a = \bd \omega^a +\frac{1}{2} \tensor{f}{^a_b_c} \omega^b\wedge\omega ^c$ is the curvature of $\omega$ and  the basis elements  $J_a$ of the Lie algebra satisfy $[J_a,J_b]=\tensor{f}{^c_a_b} J_c$, the Lie bracket of $\mathfrak{g}$. The dual basis objects $P^a$ are the generators of the inhomogeneous part of $I\mathcal{G}$. They satisfy $[P^a,P^b]=0$ and  $[P^a,J_b]=\tensor{f}{^a_b_c} P^c$. Here, $a, b, c$ run from 1 to dim $\mathfrak{g}$. The invariant metric is defined by $\langle J_a,P^b\rangle =\delta^b_a$, $\langle P^a,P^b\rangle=0=\langle J_a,J_b\rangle$.

Fortunately, consistent deformations of Chern-Simons theories have been thoroughly studied. The relevant results can be summarized in the following statement.
\begin{theorem}
\cite[page 560, eq. (14.13)]{Henneaux_Review}
\label{CS_consistent_deformations}
The consistent deformations of abelian Chern-Simons actions are Chern-Simons actions based on arbitrary groups of the same dimension.
\end{theorem}

Note that the symmetries of the internally abelianized Chern-Simons, which are $\bd \varepsilon^a$, are deformed to $\bD \varepsilon^a = \bd \varepsilon^a + \tensor{f}{^a_b_c}\omega^b \varepsilon^c$.

As we are mostly interested in gravitational theories, we would like to pay some attention to the 3-dimensional version of the Cartan-Palatini action, which is
\begin{align} \label{eq_2+1_Palatini}
    S_{\textrm{3-P}}(e, \omega) = \int_{\mathcal{M}} \tensor{e}{_i} \wedge \tensor{F}{^i} \ ,
\end{align}
where $e$ is a non-degenerate triad, $F$ is the curvature of a $SO(1, 2)$ connection $\omega$ with components $F^i=\bd \omega^i +\frac{1}{2} \tensor{\epsilon}{^i_j_k} \omega^j\wedge\omega ^k$ (here $i,j, k$ run from 1 to 3).

The internally abelianized version is obtained by replacing $SO(1, 2)$ by $U(1)^3$, so the curvature $F^i$ reduces to $\bd \omega^i$. The resulting action reads
\begin{align} \label{Palatini_3dim_abelian}
    S^0_{\textrm{3-P}}(e, \omega) = \int_{\mathcal{M}} e_i \wedge \bd \omega^i \ .
\end{align}
The field equations are
\begin{subequations}
\begin{align} \label{eq_2+1_Palatini_eom}
    \bd e^i = 0 \ , \\
    \bd \tensor{\omega}{^i} = 0 \ .
\end{align}
\end{subequations}
It is possible to give an explicit solution to them on an arbitrary manifold $\mathcal{M}$:
\begin{subequations}
\label{sol_2+1_Palatini_eom}
\begin{align} 
    \tensor{e}{^i} = \tensor{\bd f}{^i} + \tensor{\lambda}{^i _a}\tensor{\Phi}{^a} \ , \\
    \tensor{\omega}{^i} = \tensor{\bd g}{^i} + \tensor{\mu}{^i _a}\tensor{\Phi}{^a} \ ,
\end{align}
\end{subequations}
where $f^i, g^i \in \mathcal{C}^{\infty}(M)$, $\tensor{\lambda}{^i _a}, \tensor{\mu}{^i _a} \in \mathbb{R}$, and the $\tensor{\Phi}{^a}$ are representatives of the equivalence classes of $H_{dR}^1(\mathcal{M})$, the first de Rham cohomology group of $\mathcal{M}$.  Of course, if $\mathcal{M}$ is simply connected, the only terms that survive are the exact ones.
This fully parametrizes the space of solutions by $6$ functions and $2 n_1$ real numbers where $n_k = \text{dim } H_{dR}^k(\mathcal{M})$.

The degrees of freedom of the model can be interpreted and understood by looking at the pullback to the space of solutions \eqref{sol_2+1_Palatini_eom} of the presymplectic form defined through the variational principle \eqref{Palatini_3dim_abelian} (see \cite{MV} and references therein):
\begin{align*}
    \mathbf{\Omega} = - \left(\int_\Sigma \Phi^a \wedge \Phi^b\right) \ \dd \tensor{\lambda}{^i _a} \www \dd \tensor{\mu}{_i _b}  \ ,
\end{align*}
here $\Sigma$ denotes a Cauchy surface for $\mathcal{M}$. Notice that
$\Omega$ does not depend on the functions $f^i, g^i$ that we obtained previously (hence the $U(1)^6$ gauge symmetry). As a consequence, the theory has no local degrees of freedom, in fact, there are only $2 n_1$ global (``topological'') ones.

Our $U(1)^3$ Cartan-Palatini action admits many possible consistent deformations. 
Since there are six symmetries, by Theorem \ref{CS_consistent_deformations}, given any $6$-dimensional Lie group and its Lie algebra basis $A$, a consistent deformation is
\begin{align} \label{CS_deformation_formula}
    \frac{1}{3} \langle A ,[A, A] \rangle \ ,
\end{align}
with $\langle \cdot \, ,\cdot \rangle$ being an invariant metric of the corresponding Lie algebra.

It is interesting to point out that one can obtain the Cartan-Palatini action (with or without a cosmological constant) as a consistent deformation of \eqref{Palatini_3dim_abelian} by using groups closely related to $SO(2, 1)$. For instance,
consider the $\Lambda$-deformed $\mathfrak{iso}(2, 1)$ algebra
\begin{align} \label{CS_algebra}
\begin{split}
    [J_i, J_j] &= \varepsilon_{ijk} J^k \ , \\
    [J_i, P_j] &= \varepsilon_{ijk} P^k \ , \\
    [P_i, P_j] &= \Lambda \varepsilon_{ijk} J^k \ ,
\end{split}
\end{align}
equipped with the metric (that only exists in 3 dimensions)
\begin{align} \label{CS_metric}
    \langle P_i, P_j \rangle = 0 \ , \ \ \langle J_i, J_j \rangle = 0 \ , \ \ \langle P_i, J_j \rangle = \eta_{ij} \ ,
\end{align}
where $\eta$ is the Minkowski metric in 3 dimensions. This is just $\mathfrak{iso}(2, 1)$ for $\Lambda = 0$, $\mathfrak{so}(3, 1)$ for $\Lambda > 0$ and $\mathfrak{so}(2,2)$ for $\Lambda < 0$, all of them $6$-dimensional. 
A simple computation shows that, with this choice,  the deformation is given by
\begin{align*}
    S^1_{\textrm{3-P}} = \int_{\mathcal{M}} \left(\varepsilon_{ijk} e^i \wedge \omega^j \wedge \omega^k + \frac{\Lambda}{3} \varepsilon_{ijk} e^i \wedge e^j \wedge e^k\right) \ ,
\end{align*}
which, when added to the internally abelianized Cartan-Palatini action, gives back the original one \eqref{eq_2+1_Palatini} with a cosmological constant (which can be zero, positive or negative depending on the chosen group). As we can see, the three models can be obtained as consistent deformations of a very simple action with solvable field equations.

It is worth mentioning that both the 2+1 Cartan-Palatini action and its internal abelianization have the same number of local degrees of freedom (in this case zero) \cite{Witten}. This is precisely why one is a consistent deformation of the other. As we show below, this does not happen in 3+1 dimensions.

We would like to draw the attention of the readers to the fact that, according to Theorem \ref{CS_consistent_deformations} we can add any cubic interaction term that we want, as long as it has the form \eqref{CS_deformation_formula}, but one might wonder about adding \textit{quadratic terms}, which are not mentioned there and would originate in
\begin{align} \label{CS_quadratic_term}
    \langle A, \bd A \rangle \ .
\end{align}
In general, one can obtain more terms by performing a change of basis in the Lie algebra, which alters the concrete expression that we obtain, but not the objects themselves.
As a particular example, consider the action \eqref{Palatini_3dim_abelian}, that has been obtained by choosing the basis $(J_i, P_i)$ of $\mathfrak{iso}(2, 1)$ and then using $A = e^iP_i + \omega^iJ_i$.  As shown in \cite{Geille_Goeller}, one can consider instead a more general invariant metric
\begin{align} \label{CS_metric2}
    \langle P_i, P_j \rangle = \mu_1 \eta_{ij}\ , \ \ \langle J_i, J_j \rangle = \mu_2 \eta_{ij} \ , \ \ \langle P_i, J_j \rangle =  \mu_3 \eta_{ij} \ ,
\end{align}
and introduce a new basis $(J_i, T_i)$ related to the old one by $T_i = P_i + \frac{q}{2} J_i$, with $q$ constant. 
In this new basis, the brackets \eqref{CS_algebra} and metric \eqref{CS_metric2} transform into
\begin{alignat*}{3}
    &[T_i, T_j] = \varepsilon_{ijk} \left( p J^k + q T^k \right) \ , \quad
    &&[J_i, J_j] = \varepsilon_{ijk} J^k \ , \quad
    &&[J_i, T_j] = \varepsilon_{ijk} T^k \ , \\
    &\langle T_i, T_j \rangle = \sigma_1 \eta_{ij}  \ , \quad
    &&\langle J_i, J_j \rangle = \sigma_2 \eta_{ij} \ , \quad
    &&\langle J_i, T_j \rangle = \sigma_3 \eta_{ij} \ ,
\end{alignat*}
where, $p, q$, $\sigma_i$, and $\mu_i$ are related by some  expressions (irrelevant here) and can be adjusted to obtain the desired specific combinations.
Then, we write $A = \Tilde{\omega}^i J_i + \Tilde{e}^i T_i = \omega^i J_i + e^i P_i$, where
\begin{align*}
    \Tilde{\omega}^i &= \omega^i - \frac{q}{2} e^i \ , \\
    \Tilde{e}^i &= e^i \ ,
\end{align*}
achieving the effect of shifting the connection. Using the new basis in \eqref{CS_quadratic_term} and dropping the tildes, one gets the action
\begin{align} \label{MB_abelian}
    S^0 = \int_{\mathcal{M}} \left( \sigma_1 e_i \wedge \bd e^i + \sigma_2 \omega_i \wedge \bd \omega^i +2 \sigma_3 e_i \wedge \bd \omega^i \right) \ .
\end{align}
This is equivalent to \eqref{Palatini_3dim_abelian}, since we only performed a change of basis and a redefinition of the fields.
However, this is interesting because it is the internal abelianization of the Mielke-Baekler action \cite{MIELKE1991399, baekler1992dynamical}, which is the most general gravitational action in 3 dimensions:
\begin{align*}
    S_{\textrm{MB}} = \int_{\mathcal{M}} \left(  \sigma_1 e_i \wedge \bD e^i  + \sigma_2 \omega_i \wedge \left( \bd \omega^i + \frac{1}{3} [\omega, \omega]^i \right) +2 \sigma_3 e_i \wedge F^i + \frac{\sigma_4}{3} \epsilon_{ijk} e^i \wedge e^j \wedge e^k\right) \ .
\end{align*}
From the previous discussion it follows that this action admits a Chern-Simons formulation (as shown in \cite{Geille_Goeller}) which is a consistent deformation of \eqref{MB_abelian}.

Another interesting example that can be formulated as a Chern-Simons theory is the Husain model \cite{Husain}, which is a 3-dimensional version of the 4-dimensional Husain-Kucha\v{r} model. From the above discussion it follows that it is also a consistent deformation of its internally abelianized version.

\section{Deformations in four spacetime dimensions} \label{3+1}

\subsection{The Husain-Kucha\v{r} model}

The action of the Husain-Kucha\v{r} model is \cite{HK}:
\begin{align}
    S_{\textrm{HK}}(\be, \bomega) = \int_{\mathcal{M}} \tensor{\epsilon}{_i_j_k} \be^i \wedge \be^j \wedge \bF^k \ , \label{eq_HKmodel}
\end{align}
where $\bF^i$ is the curvature of a $SO(3)$ connection $\bomega^i$. This interesting model is structurally similar to General Relativity but yields no evolution ---in a concrete sense that can be understood in geometric terms--- making it a theory of 3-geometries.

Its internally abelianized version is 
\begin{align}\label{aHK}
    S^0_{\textrm{HK}}(\be, \bomega) = \int_{\mathcal{M}} \tensor{\epsilon}{_i_j_k} \be^i \wedge \be^j \wedge \bd \bomega^k \ ,
\end{align}
where $\bomega^i$ is now a $U(1)^3$ connection. 

The symmetries of $S^0_{\textrm{HK}}$ can be read directly from it. They are
\begin{subequations}\label{gsahk}
\begin{alignat}{3}
    &(Z_\rho^{0})^i_{\be} = \pounds_\rho \be^i \ &, \ \ 
    &(Z_\rho^{0})^i_{\bomega}  = \pounds_\rho \bomega^i \ , \label{gsahk1}\\
    &(Z_\tau^{0})^i_{\be}  = 0 \ &, \ \ 
    &(Z_\tau^{0})^i_{\bomega}   = \bd\tau^i \ ,\label{gsahk2}
\end{alignat}
\end{subequations}
where $\rho$ is an arbitrary vector field in $\mathcal{M}$ and the $\tau^i$ are scalar functions. These correspond to 4-diffeomorphisms and internal $U(1)$ rotations, respectively. 

The Euler forms and field equations are
\begin{subequations} \label{fehk}
\begin{alignat}{3}
    E_{\be}^{0 i} &:= -2 \tensor{\epsilon}{^i_j_k} \be^j \wedge \bd\bomega^k &= 0 \ , \\
    E_{\bomega}^{0 i} &:= -\tensor{\epsilon}{^i_j_k} \bd(\be^j \wedge \be^k) \  &= 0 \ .
\end{alignat}
\end{subequations}
In order to interpret them we first notice that the $(0,2)$-tensor $\gamma:= \be_i \otimes  \be^i$ is a degenerate metric because we only have three independent one-forms $\be_i$. If we choose a volume form $\mathsf{vol}$ in $\mathcal{M}$, we can describe the degenerate directions of the metric $\gamma$ by using
\begin{align}
{\bf{u}} (\cdot):= \left( \frac{\cdot \wedge \tensor{\epsilon}{_i_j_k}\be^i \wedge \be^j \wedge \be^k }{\mathsf{vol}}\right)\,.
\end{align}
At each point $p\in M$, this is an element of the double dual $T^{**}_p M$. As this space is canonically isomorphic to $T_p M$, ${\bf{u}}(\cdot)$ actually defines a vector field ${\bf{u}} \in\mathfrak{X}(M)$. Since $\imath_{{\bf{u}}} \be^i=  {\bf{u}} (  \be^i)=0$ we have $ \gamma ({\bf{u}}, \cdot)=0$ and, hence, the degenerate directions of the metric are those given by ${\bf{u}}$.
Also, as $\imath_{{\bf{u}}} \be^i=0$, the field equations \eqref{fehk} imply
\begin{subequations} \label{fehkcontr}
\begin{align}
   \tensor{\epsilon}{_i_j_k} \be^j \wedge \imath_{{\bf{u}}} \bd\bomega^k &= 0\,, \\
    \tensor{\epsilon}{_i_j_k} \be^j \wedge \imath_{{\bf{u}}}\bd \be^k   &= 0\,,
\end{align}
\end{subequations}
then by Lemma \ref{lemma:1form} we conclude that $\imath_{{\bf{u}}} \bd\bomega^k = 0$ and $\imath_{{\bf{u}}}\bd \be^k   = 0$. Finally, using these results we have
\begin{subequations}
\begin{align}
   \pounds_{{\bf{u}}} \bomega^i &= \imath_{{\bf{u}}} \bd\bomega^i+ \bd \imath_{{\bf{u}}} \bomega^i =  \bd  \bomega^i_{ {\bf{u}}}  \,, \\
   \pounds_{{\bf{u}}} \be^i   &=  \imath_{{\bf{u}}}\bd \be^i+  \bd\imath_{{\bf{u}}} \be^i= 0\,,
\end{align}
\end{subequations}
where $\bomega^i_{ {\bf{u}}}:= \imath_{{\bf{u}}} \bomega^i$. As we can see, the effect of Lie-dragging a solution to the field equations along the direction defined by ${\bf{u}}$ is just a gauge transformation (internal rotation) \eqref{gsahk2}. The interpretation of the dynamics from the point of view of the (degenerate) metric $\gamma$ is also clear: it will just be Lie-dragged along the integral curves of the vector field  ${\bf{u}}$.

This new model \eqref{aHK} still has the three local degrees of freedom of the full theory \eqref{eq_HKmodel}. To show this, in Appendix \ref{hamiltonian_husain_kuchar} we perform the Hamiltonian analysis of the action \eqref{aHK}. To that purpose we use the geometric implementation of Dirac’s algorithm \cite{Diracnos, HKnos}, although similar information can be obtained by using the GNH method \cite{GNH1, barbero2013, barbero2021b}.

\subsubsection{Interaction term}

The interaction term that one needs to add in order to recover the full theory is
\begin{align*}
    S^1_{\textrm{HK}}(\be, \bomega) = \int_{\mathcal{M}} \be_i \wedge \be_j \wedge \bomega^i \wedge \bomega^j \ .
\end{align*}
A direct computation gives
\begin{align*}
    \dd &S^1_{\textrm{HK}}(\mathbb{Z}_\epsilon^{0})  \\
    =& \int_{\mathcal{M}}  \bd\left( 2\tau^i \be_i \wedge \be_j \wedge \bomega^j  +  \imath_\rho \left( \be_i \wedge \be_j \wedge \bomega^i \wedge \bomega^j \right) \right) -  {\tau}^i \tensor{\varepsilon}{_i_j_k} \left( \bomega^j \wedge E_{\bomega}^{0 k}  +\be^j \wedge E_{\be}^{0 k} \right)\approx 0 \ .
\end{align*}
Hence, this deformation is consistent. We can now get the deformed symmetries by using \eqref{eq:deformation_eom}; they are
\begin{alignat*}{3}
    &(Z_\rho)^i_{\be} = \pounds_\rho \be^i \ &, \ \ 
    &(Z_\rho)^i_{\bomega} = \pounds_\rho \bomega^i \ , \\
    &(Z_\tau)^i_{\be} = -g \tensor{\varepsilon}{^i_j_k} \tau^j \be^k \ &, \ \ 
    &(Z_\tau)^i_{\bomega} = \bd\tau^i + g \tensor{\epsilon}{^i_j_k} \bomega^j \tau^k.
\end{alignat*}

\subsection{First order formulation of General Relativity} \label{sec:palatini}

The Cartan-Palatini action is
\begin{equation*}\label{HP}
S_{\textrm{P}}(\bm{\mathrm{e}},\bomega)=\int_{\mathcal{M}}\epsilon_{IJKL}\bm{\mathrm{e}}^I\wedge \bm{\mathrm{e}}^J \wedge \bF^{KL} \ ,
\end{equation*}
where $\bm{\mathrm{e}}^I$ are nondegenerate tetrads and $\bF^{IJ}$ is the curvature of a $SO(1, 3)$ connection $\bomega^{IJ}$. The indices $I, J,... = 0, 1, 2, 3$ are raised and
lowered with the Minkowski metric  $\eta_{IJ} =$ diag $(-1, +1, +1, +1)$.
Its internally abelianized version is
\begin{equation}\label{HP_free}
S^0_{\textrm{P}}(\bm{\mathrm{e}},\bomega)=\int_{\mathcal{M}}\epsilon_{IJKL}\bm{\mathrm{e}}^I\wedge \bm{\mathrm{e}}^J \wedge {\bd}{\bomega}^{KL} \ ,
\end{equation}
where now $\bomega^{IJ}$ is a $U(1)^6$ connection. As we mentioned in the introduction, this model has been studied in \cite{BarrosESa}.

The action \eqref{HP_free} gives the following Euler forms
\begin{align*}
    E_{\be}^{0 I} &= -2 \epsilon^I{}_{JKL} \be^J \wedge \bd \bomega^{KL} \ ,\\
    E_{\bomega}^{0 IJ} &= -\epsilon^{IJ}{}_{KL} \bd \left( \be^K \wedge \be^L \right) \ .
\end{align*}
It is possible to show that the field equations are equivalent to
\begin{subequations}\label{Field_eqs_free_HP}
\begin{align}
&{\bd}{\bm{\mathrm{e}}}^I=0\ ,\label{Field_eqs_HP_free1}\\
&\epsilon_{IJKL}{\bm{\mathrm{e}}}^J\wedge {\bm{\mathrm{d}\omega}}^{KL}=0\ .\label{Field_eqs_HP_free2}
\end{align}
\end{subequations}

This model is integrable as we show in the following.

\subsubsection{Solving the field equations}

In order to solve \eqref{Field_eqs_HP_free1} we use the following well known result. Let $\phi^{\scriptscriptstyle (k)}\in\Omega^k(\mathcal{M})$ be representatives of the equivalence classes in  $H^k_{\mathrm{dR}}(\mathcal{M})$ 
($i=1\,\ldots,n_k:=\mathrm{dim}\,H^k_{\mathrm{dR}}(\mathcal{M})$).  If  $\alpha\in\Omega^k(\mathcal{M})$ is such that $\bd\alpha=0$, then there exist a unique $f\in\Omega^{k-1}(\mathcal{M})$ and unique $c_i^{\scriptscriptstyle (k)}\in\mathbb{R}$ such that $\alpha=\bd f+\sum_{i=1}^{n_k}c_i^{\scriptscriptstyle (k)}\phi^{\scriptscriptstyle (k)}$. It is important to remember that the $\phi^{\scriptscriptstyle (k)}$ are closed. According to this result the solutions to \eqref{Field_eqs_HP_free1} have the form
\begin{equation}\label{sol_Field_eqs_HP_free1}
{\bm{\mathrm{e}}}^I={\bd}{\bm{\mathrm{h}}}^I+\sum_{i=1}^{n_1}{\bm{\mathrm{\eta}}}_i^I \phi^{\scriptscriptstyle (1)}_i\,,
\end{equation}
with ${\bm{\mathrm{h}}}^I\in\Omega^0(\mathcal{M})$ and ${\bm{\mathrm{\eta}}}_i^I\in \mathbb{R}$.

In order to solve \eqref{Field_eqs_HP_free2} we first notice that, as a consequence of \eqref{Field_eqs_HP_free1}, it can be written in the form ${\bd}(\epsilon_{IJKL}{\bm{\mathrm{e}}}^J\wedge {\bomega}^{KL})=0$ and, hence, we have
\begin{equation}\label{sol_Field_eqs_HP_free2}
\epsilon_{IJKL}\be^J\wedge {\bomega}^{KL}={\bd}{\bm{g}}_I+\sum_{j=1}^{n_2}{\bm{\gamma}}_{Ij}\phi^{\scriptscriptstyle (2)}_j=:B_I\,,
\end{equation}
where we have now ${\bm{g}}_I\in\Omega^1(\mathcal{M})$, ${\bm{\gamma}}_{Ij}\in\mathbb{R}$ and the closed 2-forms $\phi^{\scriptscriptstyle (2)}_j\in\Omega^2(\mathcal{M})$ are representatives of the de Rham cohomology classes in $H^2_{\mathrm{dR}}(\mathcal{M})$. Now we have to solve for ${\bomega}^{KL}$  in \eqref{sol_Field_eqs_HP_free2}. 
According to Lemma \eqref{lemma:index_eq}, we obtain
\begin{align}\label{sol_Field_eqs_HP_free2_bis}
    {\bomega}^{KL} = -\frac{1}{2} \left( \frac{B_I\wedge \be^K \wedge \be^L}{\mathsf{vol}_{\be}}\right)  \be^I -\frac{1}{2}  \left(\frac{B_I\wedge \be^I \wedge \be^{[K}}{\mathsf{vol}_{\be}} \right) \be^{L]}  \ .
\end{align}

Notice that the field equations of the full model are equivalent to Einstein's equations so, in particular, one of them states that $\bomega^{IJ}$ is the Levi-Civita spin connection. In the present abelianized example \eqref{Field_eqs_free_HP} do not imply this. 

Let us look at the simply connected case.
The metric is
\begin{align*} 
    g = \bd \bh^I \otimes \bd \bh^J \eta_{IJ} \ ,
\end{align*}
where $\bh^I$ are scalar functions. This implies that $g$ is, locally, the Minkowski metric. We see that the solutions of the model consist of all the frames that give the Minkowski metric and connections $\bomega^{IJ}$ of the form given above. Clearly, this model is not General Relativity, which ultimately explains why we have been able to solve it. 
One could hope now that deforming the action consistently would allow for more general geometries, which could lead to a nice perturbative scheme to solve Einstein's equations. Unfortunately, this is not the case.

\subsubsection{Symplectic form}
In order to disentangle the physical content of the action \eqref{HP_free} we compute the symplectic form in the solution space. This is
\begin{equation}\label{symplectic_form}
\mathbf{\Omega}_{\mathbb{S}}^\imath = \int_\Sigma \epsilon_{IJKL} \dd{\bm{\mathrm{e}}}^I\www ({\bm{\mathrm{e}}}^J\wedge \dd {\bm{\mathrm{\omega}}}^{KL}) \ .
\end{equation}
Here $\imath:\Sigma\rightarrow M$ denotes a Cauchy embedding and ${\bm{\mathrm{e}}}^I$, ${\bm{\mathrm{\omega}}}^{KL}$ denote the $\imath^*$ pullbacks to $\Sigma$ of the corresponding objects in $M$. Now, computing the variation of \eqref{sol_Field_eqs_HP_free2} we get
\[
\epsilon_{IJKL}{\bm{\mathrm{e}}}^J\wedge \dd {\bm{\mathrm{\omega}}}^{KL}=-\epsilon_{IJKL}\dd{\bm{\mathrm{e}}}^J\wedge {\bm{\mathrm{\omega}}}^{kl}+{\bd}\dd{\bm{g}}_I+\sum_{j=1}^{n_2}\dd{\bm{\gamma}}_{Ij}\phi^{(2)}_j\,,
\]
so that
\[
\dd{\be}^I\www (\epsilon_{IJKL}{\bm{\mathrm{e}}}^J\wedge \dd {\bm{\mathrm{\omega}}}^{KL})=-\epsilon_{IJKL}(\dd {\bm{\mathrm{e}}}^I\www \dd{\bm{\mathrm{e}}}^J)\wedge{\bomega}^{KL}+\dd {\bm{\mathrm{e}}}^I\www \Big( {\bd}\dd{\bm{g}}_I+\sum_{j=1}^{n_2}(\dd{\bm{\gamma}}_{Ij})\phi^{(2)}_j \Big)\,.
\]
As $\dd {\bm{\mathrm{e}}}^I\www \dd{\bm{\mathrm{e}}}^J$ is \emph{symmetric} in $IJ$ only the last term in the r.h.s. of the preceding expression contributes to $\mathbf{\Omega}_{\mathbb{S}}$. Taking into account \eqref{sol_Field_eqs_HP_free1} we see that the symplectic form in the solution space is
\begin{equation*}
\mathbf{\Omega}_{\mathbb{S}}=\int_\Sigma\Big({\bd}\dd{\bm{\mathrm{h}}}^I+\sum_{i=1}^{n_1}\dd{\bm{\mathrm{\eta}}}_i^I \phi^{\scriptscriptstyle (1)}_i\Big)\!\www\Big({\bd}\dd{\bm{g}}_I+\sum_{j=1}^{n_2}\dd{\bm{\gamma}}_{Ij}\phi^{(2)}_j\Big)\,.
\end{equation*}
As the $\phi^{(1)}_i$ and $\phi^{(2)}_j$ are closed and $\partial\Sigma=\varnothing$, the previous expression reduces to
\begin{equation}\label{Symplectic_sol_HP}
\mathbf{\Omega}_{\mathbb{S}}=\sum_{i=1}^{n_1}\sum_{j=1}^{n_2}\Big(\int_\Sigma \phi_i^{(1)}\wedge\phi_j^{(2)}\Big)\,\dd{\bm{\mathrm{\eta}}}_i^I
\,\www\, \dd {\bm{\gamma}}_{Ij} \,.
\end{equation}
As we can see the symplectic form on the solution space depends only on ${\bm{\mathrm{\eta}}}_i^I$ and ${\bm{\gamma}}_{Ij}$, i.e. a finite number of objects determined by the de Rham cohomology of $\Sigma$ (which is directly related to that of $\mathcal{M}=\mathbb{R}\times\Sigma$ by the K\"{u}nneth formula). 
Hence, the internally abelianized Cartan-Palatini action has no local degrees of freedom: the symplectic form depends on the fields only through the finite dimensional coordinates $(\bm{\mathrm{\eta}}_i^I, {\bm{\gamma}}_{Ij})$. 
This already leads to an inconsistency: the Cartan-Palatini action describes the dynamics of the two local degrees of freedom of GR, so it cannot be a consistent deformation of its internal abelianization. We will explicitly show this in the next subsection.

\subsubsection{Deformations}

In order to study the deformations, we first need to know all the symmetries of the action. Since we have the solutions given by \eqref{sol_Field_eqs_HP_free1}, \eqref{sol_Field_eqs_HP_free2}, it is straightforward to obtain the symmetry transformations of the fields.
The gauge freedom in the solutions comes from $\bm{h}_I, \bm{g}_I$, and the expressions for the symmetries of $\be^I$ and $\bomega^{IJ}$ can be directly read off from \eqref{sol_Field_eqs_HP_free1} and \eqref{sol_Field_eqs_HP_free2_bis}, respectively. 

However, there is a way to express and interpret the symmetries in a much simpler way. By defining the 2-form $\bB^I$ as in \eqref{sol_Field_eqs_HP_free2}, one can consider the invertible change of variables $(\be^I, \bomega^{IJ}) \rightarrow (\be^I, \bB^I)$.
This is equivalent to rewriting the action in the form of a BF theory as in \cite{BarrosESa}.
The main advantage of this approach is that the symmetries can be read directly and take a specially simple form, namely
\begin{alignat*}{3}
    &(Z^0_\tau)^I_{\be} = \bd \tau^I \ &, \ \ 
    &(Z^0_\tau)^I_{\bB}= 0 \ , \\
    &(Z^0_\chi)^I_{\be} = 0 \ &, \ \ 
    &(Z^0_\chi)^I_{\bB} = \bd \chi^I \ ,
\end{alignat*}
with $\tau^I$ arbitrary  functions and $\chi^I$ arbitrary 1-forms. In these expressions we readily see how internal rotations and diffeomorphisms appear on-shell (i.e. when $\bd\be^I=0$ and $\bd\bB^I=0$) as
\begin{align*}
   (Z^0_\xi)^I_{\be} = \pounds_{\xi} \be^I = \bd \left( \imath_{\xi} \be^I \right) = (Z^0_{\tau =  \imath_{\xi} \be})^I_{\be} \ , \\
    (Z^0_\xi)^I_{\bB} = \pounds_{\xi} \bB^I \approx \bd \left( \imath_{\xi} \bB^I \right) = (Z^0_{\chi =  \imath_{\xi} \bB})^I_{\bB} \ ,
\end{align*}
but the group of symmetries is larger than that of the Cartan-Palatini action.

The symmetries expressed in both sets of variables are related by
\begin{align} \label{implicit_symmetry}
    Z^I_{\bB} =  \epsilon^I{}_{JKL} \left ( Z^J_{\be} \wedge \bomega^{KL} + \be^J \wedge Z^{KL}_{\bomega} \right) \ .
\end{align}
Using Lemma \ref{lemma:index_eq}, it is immediate to find from \eqref{implicit_symmetry} that:
\begin{subequations}
\begin{align}
    (Z^0_\tau)^{IJ}_{\bomega}  =& \frac{1}{2} \tensor{\epsilon}{_K_L_P_Q} \left(\frac{\bd \tau^L \wedge \bomega^{PQ} \wedge \be^I \wedge \be^J }{\mathsf{vol}_{\be}} \right) \be^K + \frac{1}{2}  \tensor{\epsilon}{_K_L_P_Q} \left(\frac{\bd \tau^L \wedge \bomega^{PQ} \wedge \be^K \wedge \be^{[I} }{\mathsf{vol}_{\be}} \right) \be^{J]} \ , \label{symmtau} \\ 
    (Z^0_\chi)^{IJ}_{\bomega}  =& -\frac{1}{2} \left(\frac{\bd{\chi}_K\wedge \be^I \wedge \be^J}{\mathsf{vol}_{\be}} \right) \be^K -\frac{1}{2}  \left(\frac{\bd\chi_K\wedge \be^K \wedge \tensor{\be}{^[^I}}{\mathsf{vol}_{\be}} \right) \tensor{\be}{^J^]} \ .
\end{align}
\end{subequations}
Now we have to check whether the Cartan-Palatini action is a deformation of its abelian version. The interaction term that we have to add to \eqref{HP_free} is
\begin{align*}
    S^1_{\textrm{P}} = \int_{\mathcal{M}} \tensor{\epsilon}{_I_J_K_L} \be^I \wedge \be^J \wedge \tensor{\bomega}{^K_M} \wedge \bomega^{ML} \ ,
\end{align*}
which transforms as
\begin{align*}
    \dd S^1_{\textrm{P}} (\mathbb{Z}^0_\epsilon) =
    \int_{\mathcal{M}}
    2 \bd \left( \epsilon_{IJKL}\tau^I \left(  \be^J \wedge \tensor{\bomega}{^K_M} \wedge \tensor{\bomega}{^M^L} +  \be^M \wedge \tensor{\bomega}{^J_M} \wedge \tensor{\bomega}{^K^L} \right) + \chi_I \wedge \be^J \wedge \tensor{\bomega}{^I_J} \right)& \\ 
    -2 \epsilon_{IJKL}  \tau^I \bd \left( \be^J \wedge \tensor{\bomega}{^K_M} \wedge \tensor{\bomega}{^M^L} + \be^M  \wedge \tensor{\bomega}{^J_M} \wedge \tensor{\bomega}{^K^L} \right)  +2\chi_I \wedge \bd \left( \be^J \wedge \tensor{\bomega}{^I_J} \right) \neq & 0  \ ,
\end{align*}
and does not vanish on-shell because the free field equations do not imply $\bd \left( \be^J \wedge \tensor{\bomega}{^I_J} \right)=0$, as we show with a counterexample in Appendix \ref{appendix_solution}. As we can see, the full Cartan-Palatini action cannot be obtained as a consistent deformation of its internal abelianization. This means that, although this is an integrable system, we can not build upon it a regular perturbation scheme leading to GR as one would wish. 

Another interesting possibility is that one can obtain the cosmological constant term as a deformation. The interaction in this case is
\begin{align}
    S^1_{\Lambda} = \frac{\Lambda}{4!} \int_{\mathcal{M}} \tensor{\epsilon}{_I_J_K_L} \be^I \wedge \be^J \wedge \be^K \wedge \be^L \ , \label{cosmologicalterm}
\end{align}
which transforms as
\begin{align*}
    \dd S^1_{\Lambda} (\mathbb{Z}^0_\epsilon)  = \int_{\mathcal{M}} \left(  \bd \left( \frac{\Lambda}{6} \tensor{\epsilon}{_I_J_K_L} \tau^I \be^J \wedge \be^K \wedge \be^L  \right) - \frac{\Lambda}{4}  \tau^I \be^J \wedge E^0_{\bomega_{IJ}} \right) \approx 0 \ ,
\end{align*}
and is indeed a consistent deformation. Notice that the last term implies a deformation in the symmetry \eqref{symmtau} by $-g \frac{\Lambda}{8}  \left(\tau^I \be^J - \tau^J \be^I \right)$.

\subsection{General Relativity from the Holst action}

The Holst action \cite{Holst:1995pc} is
\begin{equation*}
S_{\textrm{H}}(\bm{\mathrm{e}},\bomega)=\int_{\mathcal{M}}P_{IJKL}\bm{\mathrm{e}}^I\wedge \bm{\mathrm{e}}^J \wedge \bF^{KL} \ ,
\end{equation*}
where $\bF^{IJ}$ is the curvature of an $SO(1, 3)$ connection $\bomega^{IJ}$ (with $\bomega^{IJ}=-\bomega^{JI}$) and
\begin{align*}
    P_{IJKL}:=\frac{1}{2}\left(\epsilon_{IJKL}-\frac{1}{\gamma}\eta_{IK}\eta_{JL}+\frac{1}{\gamma}\eta_{JK}\eta_{IL}\right) \ .
\end{align*}
Its internally abelianized version is
\begin{equation}\label{Holst_free}
S^0_{\textrm{H}}(\bm{\mathrm{e}},\bomega)=\int_{\mathcal{M}}P_{IJKL}\bm{\mathrm{e}}^I\wedge \bm{\mathrm{e}}^J \wedge {\bd}{\bomega}^{KL} \ ,
\end{equation}
where now $\bomega^{IJ}$ is a $U(1)^6$ connection (the 6 comes from the fact that the $\bomega^{IJ}$ are six independent 1-forms).

The field equations derived from \eqref{Holst_free} are equivalent to
\begin{subequations}\label{Field_eqs_free_Holst}
\begin{align}
&\bd {\bm{\mathrm{e}}}^I=0\,,\label{Field_eqs_Holst_free1}\\
&P_{IJKL}{\bm{\mathrm{e}}}^J\wedge {\bm{\mathrm{d}\omega}}^{KL}=0\,.\label{Field_eqs_Holst_free2}
\end{align}
\end{subequations}
A couple of comments are in order now. First, we point out that equation \eqref{Field_eqs_Holst_free1} is obtained in the same way both for the Holst and the Cartan-Palatini actions. 
The second is that, at variance with the full actions, the field equations for their free counterparts \emph{are not equivalent}. This is because there is no reason,  \emph{a priori}, to expect that ${\bm{\mathrm{e}}}^I\wedge{\bm{\mathrm{d}\omega}}_I^{\phantom{I}J}=0$ (remember that $\bD {\bm{\mathrm{e}}}^I=0$ and the Bianchi identity imply ${\bm{\mathrm{e}}}^I\wedge {\bm{\mathrm{F}}}_I^{\phantom{I}J}=0$). As a consequence, the deformation analysis of the Holst action is different from the Cartan-Palatini one. In any case, the internally abelianized version of the Holst action also has no local degrees of freedom, as can be seen by computing the presymplectic form in the solution space, which only depends on a finite number of cohomological variables (we omit the details since the argument is essentially the same as the one discussed above for the internally abelianized Cartan-Palatini action). We then conclude that the Holst action cannot be obtained as a consistent deformation of its internal abelianization.

In spite of this, it is still worth studying the `pure Holst' term separately
\begin{align*}
    S_{\textrm{pH}} = \int_{\mathcal{M}} \be_I \wedge \be_J \wedge \bF^{IJ} \ .
\end{align*}
This action was discussed in \cite{Montesinos:Holst} and found to be topological.  
Its internal abelianization is
\begin{align*}
    S^0_{\textrm{pH}} =  \int_{\mathcal{M}} \be_I \wedge \be_J \wedge \bd \bomega^{IJ} \ ,
\end{align*}
with Euler forms
\begin{alignat*}{3}
    E_{\be}^{0 I} &=- 2 \be_J \wedge \bd \bomega^{IJ} \ , \\
    E_{\bomega}^{0 IJ} &=- \bd  \left( \be^I \wedge \be^J \right) \ ,
\end{alignat*}
leading to the field equations
\begin{align*}
    \be_J \wedge \bd \bomega^{IJ} \ &= 0 \ ,\\
    \bd \be^I &= 0 \ .
\end{align*}
This action again admits a BF formulation if one defines
\begin{align*}
    \bB^I := \be_J \wedge \bomega^{IJ} \ ,
\end{align*}
which leads to the action
\begin{align*}
    S^0_{\textrm{pH}} = -2\int_{\mathcal{M}} \bd \be_I \wedge \bB^I \ . 
\end{align*}
With these variables, the symmetries for $\be, \bB$ are the same as in the Cartan-Palatini case and those for $\bomega$ are implicitly given by
\begin{align*}
    Z^I_{\bB} =  Z^J_{\be} \wedge \tensor{\bomega}{^I_J} + \be^J \wedge \tensor{Z}{_\bomega^I_J} \ ,
\end{align*}
which yield the symmetry transformations of $\bomega$ (see lemma \ref{lemma:last_equation})
\begin{align*}
    {(Z^0_\tau)_{\bomega}}_{IJ} &= \frac{1}{2} \epsilon_{PQK[I} \left( \frac{\bomega_{J]M} \wedge \bd\tau^M  \wedge \be^P \wedge \be^Q}{\mathsf{vol}_{\be}} \right) \be^K - \frac{1}{4} \epsilon_{IJPQ} \left( \frac{\bd\tau^M \wedge \bomega_{MK} \wedge \be^P \wedge \be^Q}{\mathsf{vol}_{\be}} \right) \be^K , \\ 
    {(Z^0_\chi)_{\bomega}}_{IJ} &= \frac{1}{2} \epsilon_{PQK[I} \left( \frac{\bd\chi_{J]} \wedge \be^P \wedge \be^Q}{\mathsf{vol}_{\be}} \right) \be^K - \frac{1}{4} \epsilon_{IJPQ} \left( \frac{\bd\chi_K \wedge \be^P \wedge \be^Q}{\mathsf{vol}_{\be}} \right) \be^K \ .
\end{align*}
The interaction term that we need to recover the full pure Holst term is
\begin{align*}
    S^1_{\textrm{pH}} = \int_{\mathcal{M}} \be_I \wedge \be_J \wedge \tensor{\bomega}{^I^K} \wedge \tensor{\bomega}{_K^J} =  \int_{\mathcal{M}} \bB_K \wedge \bB^K\,. 
\end{align*}
We have now
\begin{align*}
    \dd S^1_{\textrm{pH}} (\mathbb{Z}^0_\epsilon) &= \int_{\mathcal{M}} \bigg( 2 \bd \left( \chi_K \wedge \be^J \wedge \tensor{\bomega}{^K_J}  \right)  + \chi_K \wedge E_{\be}^{0 K}  \\ 
    &\hspace*{-12mm}\left. + \left(\frac{1}{4} \epsilon_{IJPQ} \left( \frac{\chi_R \wedge \tensor{\bomega}{_K^R} \wedge \be^P \wedge \be^Q}{\mathsf{vol}_{\be}} \right)- \frac{1}{2} \epsilon_{PQK[I} \left( \frac{\chi_R \wedge \tensor{\bomega}{_{J]} ^R} \wedge \be^P \wedge \be^Q}{\mathsf{vol}_{\be}} \right)     \right)\be^K \wedge E_{\bomega}^{0 IJ}  
    \right)  \approx 0 \ .
\end{align*}
Hence, the deformed symmetries are
\begin{align*}
    &(Z_\tau)_{\be}^I = \bd \tau^I \ , \\
    &{(Z_\tau)_{\bomega}}_{IJ} = \frac{1}{2} \epsilon_{PQK[I} \left( \frac{\bomega_{J]M} \wedge \bd\tau^M  \wedge \be^P \wedge \be^Q}{\mathsf{vol}_{\be}} \right) \be^K - \frac{1}{4} \epsilon_{IJPQ} \left( \frac{\bd\tau^M \wedge \bomega_{MK} \wedge \be^P \wedge \be^Q}{\mathsf{vol}_{\be}} \right) \be^K \ , \\
    &(Z_\chi)_{\be}^I =  g \chi^I \ , \\ 
    &{(Z_\chi)_{\bomega}}_{IJ} = \frac{1}{2} \epsilon_{PQK[I} \left( \frac{\bd\chi_{J]} \wedge \be^P \wedge \be^Q}{\mathsf{vol}_{\be}} \right) \be^K - \frac{1}{4} \epsilon_{IJPQ} \left( \frac{\bd\chi_K \wedge \be^P \wedge \be^Q}{\mathsf{vol}_{\be}} \right) \be^K  \\
    &\quad\quad  + g\frac{1}{4} \epsilon_{IJPQ} \left( \frac{\chi_R \wedge \tensor{\bomega}{_K^R} \wedge \be^P \wedge \be^Q}{\mathsf{vol}_{\be}} \right) \be^K  -g\frac{1}{2} \epsilon_{PQK[I} \left( \frac{\chi_R \wedge \tensor{\bomega}{_{J]} ^R} \wedge \be^P \wedge \be^Q}{\mathsf{vol}_{\be}} \right)\be^K \ .
\end{align*}
The intricate structure of these gauge transformations is quite remarkable and, in fact, difficult to guess just by simply looking at the pure Holst action. Notice that this action is a deformation of its internal abelianization, which is consistent since both models have no local degrees of freedom. It can also be shown that as in the internally abelianized Cartan-Palatini case, the cosmological constant term \eqref{cosmologicalterm} is indeed a consistent deformation of the internally abelianized Holst  and ``pure Holst'' actions.

Let us remark, as a conclusion, that the problematic piece is the Cartan-Palatini action, with or without the pure Holst term, since it is this term that ``loses'' local degrees of freedom when we take its internal abelianization.

\subsection{The self-dual action}

In \cite{Smolin}, Smolin considers the internally abelianized self-dual action (essentially the Cartan-Palatini action written in terms of self-dual connections satisfying $\star\bomega = \bomega$, where $\star$ is the internal Hodge dual). Since in 4 dimensions $\star^2 = -\text{Id}$, this operator is diagonalizable in the complex numbers, yielding
\begin{align*}
    \star = i P_+ - i P_- \ ,
\end{align*}
where $P_+, P_-$ are projectors onto the self and anti-self dual subspaces, respectively. Introducing indices, one writes
\begin{align*}
    \tensor{P}{_+^I^J_K_L} = \frac{1}{2} \left( \tensor{\eta}{^I_K}\tensor{\eta}{^J_L} - \tensor{\eta}{^J_K} \tensor{\eta}{^I_L} - i\tensor{\epsilon}{^I^J_K_L}  \right) = - i \tensor{P}{_{\gamma = -i}^I^J_K_L} \ ,
\end{align*}
which is a multiple of Holst's $P$ with $\gamma = -i$ (recall that in the Lorentzian case, $P$ is not invertible if $\gamma = \pm i$).

Then, the action can be written in the convenient form
\begin{align} 
\label{selfdual_Palatini}
    S^0_{\textrm{sdP}} \left( \be, \bomega\right) = \int_{\mathcal{M}} \tensor{\epsilon}{_I_J_K_L} \be^I \wedge \be^J \wedge \bd  \left( \tensor{P}{_+^K^L_M_N}\bomega^{MN} \right) \ .
\end{align}

Notice that we could have omitted the projector from the action and just have considered self-dual connections. In that case, the variations of the connection must be self-dual themselves.
In \eqref{selfdual_Palatini}, we are considering arbitrary connections since the projector already takes care of self-duality (both of the fields and their variations).

If one only considers real fields, then one can split the action into a real and an imaginary part
\begin{align*}
    S^0_{\textrm{sdP}} \left( \be, \bomega\right) = 
    \int_{\mathcal{M}} \tensor{\epsilon}{_I_J_K_L} \be^I \wedge \be^J \wedge \bd  \bomega^{KL}
    - 2i\int_{\mathcal{M}} \be_I \wedge \be_J \wedge \bd  \bomega^{IJ} \ ,
\end{align*}
both of which must satisfy the stationary principle \textit{separately}. This gives the equations of motion for the Palatini \textit{and} pure Holst actions.

On the other hand, if one works with complex fields as Smolin suggests in \cite{Smolin}, there are a few details to take into account.
Since the metric must be real, one has to additionally impose by hand the so-called reality conditions to enforce this condition. This can be done by introducing second-class constraints \cite{Morales}, but they do not come out naturally from the formalism, rather they are a supplement to the action.
One should also realize that the mathematical structures involved in the canonical analysis (such as the symplectic form) change in the passage from the real to the complex case.

One could also argue about the relevance of the complex field formulation in the current landscape. Smolin's analysis precedes the forgoing of complex fields in favor of real fields, which partially clarified some issues and were quickly adopted. However, it has been recently suggested that complex fields might still be relevant \cite{Perez, Wieland}.
In view of all this, it is of our opinion that one should carefully repeat the Hamiltonian analysis of the complex self-dual action from scratch from a modern point of view, for instance, along the lines put forward in \cite{BDMV}.

It is well-known that in the Euclidean case the self-dual action can be written in terms of real fields. A neat way to do it was proposed in \cite{Barbero_1995}. The action is
\begin{align} \label{eq_asd_action}
    S_{\textrm{asd}}(\be, \balpha, \bomega) = \int_{\mathcal{M}} \left( \epsilon_{ijk} \be^i \wedge \be^j \wedge \bF^k - 2 \balpha \wedge \be_i \wedge \bF^i \right) \ ,
\end{align}
where $\bF^i$ is the curvature of a $SO(3)$ connection  $\bomega^i$ and $\balpha$ is essentially $\be^0$. This comes from considering a Palatini-like action with anti-self-dual connections and writing it in terms of three one-forms $\bomega^i$. This model has 2 local degrees of freedom.

The internal abelianization of \eqref{eq_asd_action} yields
\begin{align}
\label{eq_action_abelian_Fernando}
    S^0_{\textrm{asd}}(\be, \balpha, \bomega) = \int_{\mathcal{M}} \left( \epsilon_{ijk} \be^i \wedge \be^j \wedge \bd \bomega^k - 2 \balpha \wedge \be_i \wedge \bd \bomega^i \right) \ ,
\end{align}
with Euler forms
\begin{align*}
    E^0_{\balpha} &= 2 \be_i \wedge \bd \bomega^i \ , \\
    {E^0_{\be}}_i &= -2 \left( \epsilon_{ijk}\be^j \wedge \bd \bomega^k + \balpha \wedge \bd \bomega_i \right) \ , \\
    {E^0_{\bomega}}_k &=  -\bd \left( \epsilon_{ijk} \be^i \wedge \be^j - 2 \balpha \wedge \be_k \right) \ .
\end{align*}
The symmetries of the internally abelianized model are
\begin{alignat*}{3}
    &{(Z^0_\tau)_{\bomega}}^{i} = \bd \tau^i \ , \quad
    &&{(Z^0_\Lambda)_{\bomega}}^{i} = 0 \ , \quad
    &&{(Z^0_\xi)_{\bomega}}^{i} = \pounds_{\xi} \bomega^i\ , \\
    &{(Z^0_\tau)_{\be}}^{i} = 0 \ , \quad 
    &&{(Z^0_\Lambda)_{\be}}^{i} = - \Lambda^i \balpha + \tensor{\epsilon}{^i_ j_k} \be^j \Lambda^k  \ ,  \quad
    &&{(Z^0_\xi)_{\be}}^{i} = \pounds_{\xi} \be^i \ , \\ 
    &(Z^0_\tau)_{\balpha}= 0 \ ,  \quad
    &&(Z^0_\Lambda)_{\balpha} = \Lambda_i \be^i  \ ,  \quad
    &&(Z^0_\xi)_{\balpha}= \pounds_{\xi} \balpha \ .
\end{alignat*}

As in the case of Husain-Kucha\v{r} detailed in Appendix \ref{hamiltonian_husain_kuchar}, it is possible to get the Hamiltonian description of \eqref{eq_action_abelian_Fernando} by following the procedure discussed in \cite{Barbero_1995}. One finds that this abelianized model still has two local degrees of freedom, hence there is hope that the full version can be obtained as a consistent deformation.

The interaction term needed to recover the full action is
\begin{align*}
    S^1_{\textrm{asd}} = \int_{\mathcal{M}} \left( \be_i \wedge \be_j \wedge \bomega^i \wedge \bomega^j - \epsilon_{ijk} \balpha \wedge \be^i \wedge \bomega^j \wedge \bomega^k \right) \ .
\end{align*}
Under a symmetry transformation it changes as
\begin{align*}
    \dd S^1_{\textrm{asd}} \left( \mathbb{Z}^0_\epsilon \right) &= \int_{\mathcal{M}} \left(  2\bd \left( \tau^i  \left(\be_i \wedge \be_j \wedge \bomega^j + \epsilon_{ijk} \balpha \wedge \be^j \wedge \bomega^k \right)\right) \right. \\
    & + \bd \imath_{\xi} \left( \be_i \wedge \be_j \wedge \bomega^i \wedge \bomega^j - \epsilon_{ijk} \balpha \wedge \be^i \wedge \bomega^j \wedge \bomega^k \right)  \\
    &\left. +\tau^i \left( -\epsilon_{ijk} \bomega^j \wedge E_{\bomega}^{0 k} - \be_i \wedge E^0_{\balpha} + \balpha \wedge {E^0_{\be}}_i \right) \right) \approx 0 \ ,
\end{align*}
hence it is a consistent deformation. The only symmetry that changes with respect to the internally abelianized model is
\begin{align*}
    &{(Z_\tau)_{\bomega}}^{i} = \bd \tau^i +  g\tensor{\epsilon}{^i_j_k} \bomega^j \tau^k = \bD \tau^i \ , \\ 
    &{(Z_\tau)_{\be}}^{i} = g\tau^i \balpha \ , \\ 
    &{(Z_\tau)_{\balpha}} = - g\tau^i \be_i \ .
\end{align*}
One can easily check that these are the symmetries found in \cite{Barbero_1995} by taking linear combinations. Furthermore, the cosmological constant term \eqref{cosmologicalterm} is also a consistent deformation of \eqref{eq_action_abelian_Fernando}.

The previous result is rather interesting, since it shows how General Relativity (Euclidean, though) can \textit{actually} be obtained as a perturbation of a simpler abelianized model, in contrast with the situation with the Cartan-Palatini action. Notice that the very same approach would not work for Lorentzian General Relativity, as using anti-self-dual fields in that case forces them to be complex. At any rate, the possibility that a real action similar to \eqref{eq_asd_action} giving Lorentzian GR exists cannot be ruled out \textit{a priori}. Finding it would open the possibility of a sensible perturbative approach.

\section{Conclusions}

Inspired by the ideas proposed in \cite{Smolin} by Smolin, we have studied the internal abelianization and consistent deformations of several of the most relevant gravitational theories in three and four spacetime dimensions. We would like to remark that we have treated the problem in a fully covariant way; no foliations or time specifications have been used.  A similar treatment would be possible in the Hamiltonian framework, however it would be less directly connected to the action and its symmetries and more tedious. One could indeed try the strategy that we have followed here within the canonical approach but, ultimately, the results would be the same.

In the three-dimensional case, we have seen how consistent deformations work for Chern-Simons theories and we have considered in detail the Cartan-Palatini action, which is a particular case. We have seen that the internal abelianization is integrable and we have given its solution. Moreover, there exist multiple consistent deformations, each with its own interpretation. Hence, it is possible to pursue a standard perturbative approach, in particular leading to full 2+1 gravity. It is important to remark, however, that it is possible to obtain many more actions from the Chern-Simons one, in particular gravitational models with torsion and nonmetricity, and even supersymmetric extensions \cite{Cacciatori, Giacomini}. This strongly suggests that all gravitational theories in three dimensions can be derived from a Chern-Simons action.

In the four-dimensional case, we have shown that the Husain-Kucha\v{r}, pure Holst and real anti-self-dual (Euclidean gravity) actions are consistent deformations of their internally abelianized versions. Only in the Cartan-Palatini and full Holst cases, which are the physically most interesting ones, does the approach fail and the perturbation becomes singular.
In all the other actions analyzed, the number of degrees of freedom is conserved, however for these particular cases, the internally abelianized theory is only topological, while General Relativity has two local degrees of freedom. 
Since the full Cartan-Palatini and Holst theories have the same symmetries, the Holst action can be seen as a deformation of the Cartan-Palatini action.  In this sense, the only link that is missing is, precisely, the one that would allow for a regular perturbative treatment. Although this is somewhat disappointing, this does not mean that the problem cannot be treated perturbatively, just that the standard asymptotic expansion approach does not work. One can also make similar checks in the real BF-type \cite{Capovilla, Livine, Engle} and MacDowell-Mansouri \cite{Macdowell1977} approaches by using internally abelianized versions equivalent to those already studied in the previous sections. 

It is important to highlight the difference between the standard perturbative approach and internal abelianization. In the first case, one usually picks a fixed background with a clear physical interpretation (a particular solution to the field equations with relevant symmetries), introduces dynamical fields as perturbations of the background and, finally, expands the action in terms of these perturbations. This way one usually obtains a quadratic (``free'') theory and interaction terms. Abelianization, on the other hand, means that the internal gauge group is replaced by an Abelian group with the same dimension. This has some important implications. First,  the interaction terms that disappear from the Lagrangian are not the same ones eliminated in the usual perturbative approaches. Second, some important symmetries, such as diff-invariance, can be kept in the process. And finally, and related to the previous comment, no background objects are introduced. Notice, also, that any field theory admits a perturbative treatment whereas internal abelianization only makes sense for some types of gauge theories.

Free theories are a natural starting point to build interacting models by studying their consistent deformations. In fact, this has been done for Yang-Mills theories \cite{Henneaux} (starting from several copies of the Maxwell action) and general relativity \cite{Boulanger2001} (starting from the Fierz-Pauli Lagrangian). It is interesting to understand what happens if both approaches are combined. Specifically, take first an expansion to second order of the action and then abelianize the internal symmetry group. This procedure could be simpler than directly studying the consistent perturbations of the abelianized action and could provide a quick approach to the problem. In Appendix \ref{lacp} we study the concrete example provided by the Cartan-Palatini action and show that the simplified model has no local degrees of freedom so, the Cartan-Palatini action is not a consistent deformation of it. In any case we would like to emphasize that the procedure that we give in the paper is quite transparent and of general applicability, hence, if no \textit{ad hoc} shortcuts can be found for a particular model one can always rely on it. 

An interesting idea for future work would be to study perturbations of the Husain-Kucha\v{r} model, which although not describing General Relativity, is closely related to it, as it is a theory of 3-geometries. 
Another interesting line on work would be based on the anti-self-dual action. One could hope that there exists a similar action which still is a consistent deformation of its internally abelianized version but leads to Lorentzian General Relativity.

\section*
{Acknowledgments}
We thank the referee for some interesting comments that have helped us improve the paper.
This work has been supported by the Spanish Ministerio de Ciencia Innovaci\'on y Uni\-ver\-si\-da\-des-Agencia Estatal de Investigaci\'on PID2020-116567GB-C22 grants. E.J.S. Villase\~nor is supported by the Madrid Government (Comunidad de Madrid-Spain) under the Multiannual Agreement with UC3M in the line of Excellence of University Professors (EPUC3M23), and in the context of the V PRICIT (Regional Programme of Research and Technological Innovation). Bogar D\'iaz acknowledges support from the CONEX-Plus programme funded by Universidad Carlos III de Madrid and the European Union's Horizon 2020 research and innovation programme under the Marie Sk{\l}odowska-Curie grant agreement No. 801538. 

\appendix
\section{A few useful lemmas}

\begin{lemma}\label{lemma:Lms}
Let $\Sigma$ be an orientable 3-dimensional manifold. For given $w_i\in\Omega^2(\Sigma)$ and $e^i\in\Omega^1(\Sigma)$ a nondegenerate coframe, with volume form $\bm \omega:= \frac{1}{3!} \epsilon_{ijk} e^i \wedge e^j \wedge e^k$,
consider the following system of equations in the unknowns $v^i\in\Omega^1(\Sigma)$
\begin{equation*}\label{equations}
  \epsilon_{ijk}e^j\wedge v^k=w_i\,.
\end{equation*}
Then the solution is
\begin{equation*}\label{solution}
v^i=\left(\frac{e^j\wedge w_j}{2 \ \bm \omega}\right) e^i- \left(\frac{e^i\wedge w_j}{\bm \omega}\right) e^j\,.
\end{equation*}
\end{lemma}

\begin{lemma}
\label{lemma:1form}
    Let $\mathcal{M}$ be a 4-dimensional parallelizable manifold, $\be^i\in\Omega^1(\mathcal{M})$ with $i=1,2,3$ linearly independent 1-forms and $\mathbf{S}^i\in\Omega^1(\mathcal{M})$ be another three 1-forms. Then, $\tensor{\varepsilon}{_i_j_k}\be^j\wedge \mathbf{S}^k=0$ implies $\mathbf{S}^i=0$.
\end{lemma}

\begin{lemma}
Let $\mathcal{M}$ be a 4-dimensional parallelizable manifold and $\be^I \in\Omega^1(\mathcal{M})$ a nondegenerate coframe.
Any $2-$form $\mathbf{S} \in \Omega^2(\mathcal{M})$ can be written as
\begin{align*}
    \mathbf{S} = \frac{1}{4} \epsilon_{JKLM} \left(  \frac{\mathbf{S}^I \wedge \be^L \wedge \be^M}{\mathsf{vol}_{\be}}\right) \be^J \wedge \be^K \ . 
\end{align*}
\end{lemma}

\begin{lemma}
\label{lemma:index_eq}
    Let $\mathcal{M}$ be a 4-dimensional parallelizable manifold and $\be^I \in\Omega^1(\mathcal{M})$ a nondegenerate coframe.
    The solution to
    \begin{align*}
        \epsilon_{IJKL} \be^J \wedge Z^{KL} = \Phi_I \ ,
    \end{align*}
    for $Z^{KL}\in\Omega^1(\mathcal{M})$ antisymmetric in $KL$ and $\Phi_I\in \Omega^2(\mathcal{M})$ is 
    \begin{align*}
        Z^{KL} = -\frac{1}{2} \left( \frac{\Phi_I \wedge \be^K \wedge \be^L}{\mathsf{vol}_{\be}} \right) \be^I -\frac{1}{2} \left( \frac{\Phi_I \wedge \be^I \wedge \be^{[K}}{\mathsf{vol}_{\be}} \right) \be^{L]} \ .
    \end{align*}
\end{lemma}

\begin{lemma}
\label{lemma:last_equation}
    Let $\mathcal{M}$ be a 4-dimensional parallelizable manifold and $\be^I \in\Omega^1(\mathcal{M})$ a nondegenerate coframe.
    The solution to the equation 
    \begin{align*}
        \be^J \wedge Z_{IJ} = \Phi_I \ ,
    \end{align*}
    for $Z_{IJ}\in\Omega^1(\mathcal{M})$ antisymmetric in $IJ$ and $\Phi_I\in \Omega^2(\mathcal{M})$ is 
    \begin{align*}
        Z_{IJ} = \frac{1}{2} \epsilon_{PQK[I} \left( \frac{\Phi_{J]} \wedge \be^P \wedge \be^Q}{\mathsf{vol}_{\be}} \right) \be^K - \frac{1}{4} \epsilon_{IJPQ} \left( \frac{\Phi_K \wedge \be^P \wedge \be^Q}{\mathsf{vol}_{\be}} \right) \be^K\ .
    \end{align*}
\end{lemma}
The volume form appearing in the preceding lemmas is the one defined in \eqref{vole}.

\section{An interesting example} \label{deformation_symmetry}

It is worth taking a moment to reflect on the fact that being a consistent deformation of an action is not a symmetric property.
Consider the actions
\begin{align*}
    S_{3\textrm{-P}}(e, \omega) = \int_{\mathcal{M} } e_i \wedge F^i \ , \ \ \ S_{YM}(\omega) = \int_{\mathcal{M}} F_i \wedge \star F^i \ .
\end{align*}
In 3 dimensions, general relativity (as given by the above Cartan-Palatini action) has no local degrees of freedom, while Yang-Mills has $1\times \textrm{dim}\, SO(3)= 1 \times 3=3$. 
If we consider the sum of both actions we obtain a new one whose equations of motion are
\begin{align*}
   2\ \bD \star F^i + \bD e^i = 0 \ , \\
    F^i = 0 \ ,
\end{align*}
which are trivially equivalent to
\begin{align*}
    \bD e^i = 0 \ , \\
    F^i = 0 \ ,
\end{align*}
which, in turn, are the equations of General Relativity in 3 dimensions, a theory with no local degrees of freedom. As we see, if we start with the Cartan-Palatini action and add the Yang-Mills term as an interaction, this is a consistent deformation. On the other hand, if the starting point is Yang-Mills and we add the gravitational term as an interaction the initial and final numbers of degrees of freedom do not match and, hence, we do not have a consistent deformation.

\section{Hamiltonian analysis of the internally abelianized Husain-Kucha\v{r} model}
\label{hamiltonian_husain_kuchar}
The Lagrangian defined by the action \eqref{aHK} after performing a 3+1 decomposition is
\begin{align*}\begin{split}
L({\mathrm{v}} )\!=\!\int_{\Sigma} & \epsilon_{ijk} \Big( \!\left( v_\omega^{i}- \mathrm{d} \omega^i_{\mathrm{t}} \right)\!\wedge\!  e^{j} \!\wedge\! e^{k}+ 2e^i_{\mathrm{t}} e^{j}\!\wedge\! \mathrm{d} \omega^{k}  \Big)\,,\end{split}
\end{align*}
where $\mathrm{d}$ denotes the exterior differential on $\Sigma$, the variables $\omega^i$ and $e^i$ are now an $U(1)^3$-connection and a frame field on $\Sigma$ respectively, the $\omega_{\mathrm{t}}^i, e_{\mathrm{t}}^i \in C^\infty (\Sigma)$ are scalar fields and
\[
\mathrm{v}:=((\omega_{\mathrm{t}}^i,\omega^i,e_{\mathrm{t}}^i,e^i),(v_{\omega \mathrm{t}}^i,v_\omega^i,v_{e \mathrm{t}}^i,v_e^i))
\]
denotes a point of the tangent bundle $T\mathcal{Q}$  of  the configuration space $\mathcal{Q}$ (defined by $(\omega_{\mathrm{t}}^i,\omega^i,e_{\mathrm{t}}^i,e^i)$). If we take $\mathrm{v}, \mathrm{w}$ in the same fiber of $T\mathcal{Q}$,
we obtain the fiber derivative
\begin{align*}
&\left\langle F\!L \left(\mathrm{v}\right)|\mathrm{w}\right\rangle=  \int_{\Sigma} \epsilon_{ijk}  w_\omega^{i}\wedge e^{j} \wedge e^{k} \,.
\end{align*}
This implies that we have the following primary constraints
\begin{subequations}\label{consHKpb}
\begin{align}
{\bf C}_{\mathrm{t} i}(\cdot)&:= {\bf P}_{\!\!\mathrm{t} i}(\cdot) = 0 \,, & \hspace*{-3mm} {\bf C}_{i}(\cdot)&:=  \displaystyle {\bf P}_{i}(\cdot)-\!\!\int_{\Sigma} \!\cdot \wedge  \epsilon_{ijk} e^{j} \wedge e^{k}=0\,,\\
{\bf c}_{\mathrm{t} i}(\cdot)&:= {\bf p}_{\mathrm{t} i}(\cdot) = 0     \,, &  \hspace*{-3mm} {\bf c}_{i}(\cdot)&:= {\bf p}_{i}(\cdot) = 0\,,
\end{align}
\end{subequations}
where here and in the following the points $(q,{\bf p})\in T^*\mathcal{Q}$ will be denoted as
\begin{equation*}
 (q,{\bf p}):=\big((\omega_{\mathrm{t}}^i,\omega^i,e_{\mathrm{t}}^i,e^i),({\bf P}_{\!\!\mathrm{t} i},{\bf P}_{ i},{\bf p}_{\mathrm{t} i},{\bf p}_{i})\big)\,.
\end{equation*}
Notice that ${\bf C}_{\mathrm{t} i}(\cdot)$ and ${\bf c}_{\mathrm{t} i}(\cdot)$ are linear functionals acting on $\mathfrak{u}(1)^3$-valued scalar functions on $\Sigma$ whereas ${\bf C}_{i}(\cdot)$ and ${\bf c}_{i}(\cdot)$ are linear functionals acting on $\mathfrak{u}(1)^3$-valued 1-forms.

The Hamiltonian is only defined on the primary constraint submanifold. An extension of it to the full phase space of the model is
\begin{align*}
H=\int_{\Sigma} &  \epsilon_{ijk}\left( \mathrm{d} \omega^i_{\mathrm{t}}\wedge  e^{j}\wedge e^{k}- 2e^i_{\mathrm{t}} e^{j}\wedge \mathrm{d} \omega^{k}\right)\,.
\end{align*}
By writing tangent vectors $Z\in T_{(q,{\bf p})}T^*\mathcal{Q}$ as
\begin{align*}
  Z&:=\left((q,{\bf p}),(Z^i_{\omega_{\mathrm{t}}},Z^i_{\omega},Z^i_{e {\mathrm{t}}},Z^i_{e}, {\bf Z}_{{\bf P} {\mathrm{t}} i},{\bf Z}_{{\bf P} i}, {\bf Z}_{{\bf p} {\mathrm{t}} i},{\bf Z}_{{\bf p} i})\right)\,,
\end{align*}
the canonical symplectic form $\Omega$ acting on vector fields on $T^*\mathcal{Q}$ becomes
\begin{align*}
\Omega(X,Y)=&{\bf Y}_{{\bf P} {\mathrm{t}} i}\left(X^i_{\omega {\mathrm{t}} }\right)-{\bf X}_{{\bf P} {\mathrm{t}}i}\left(Y^i_{\omega {\mathrm{t}}}\right)+{\bf Y}_{{\bf P}i}\left(X^i_{\omega}\right)-{\bf X}_{{\bf P} i}\left(Y^i_{\omega}\right)\nonumber\\
+&{\bf Y}_{{\bf p} {\mathrm{t}} i}\left(X^i_{e {\mathrm{t}}}\right)\,-{\bf X}_{{\bf p} {\mathrm{t}} i}\left(Y^i_{e {\mathrm{t}}}\right)\,+{\bf Y}_{{\bf p} i}\left(X^i_{e}\right)\,-{\bf X}_{{\bf p} i}\left(Y^i_{e}\right)\,.
\end{align*}
The implementation of the geometric form of the Dirac algorithm described in \cite{BDMV} is now a straightforward exercise. The main step is solving for the Hamiltonian vector field $X$ in the equation
\begin{align*}
\Omega(X,Y)&= \, d H \left(Y\right) +\langle \lambda^i_{\mathrm{t}} |d {\bf C}_{{\mathrm{t}} i}\rangle \left(Y\right)+\langle \lambda^i |d {\bf C}_{i}\rangle \left(Y\right)+  \langle  \mu^i_{\mathrm{t}}| d {\bf c}_{{\mathrm{t}} i} \rangle \left(Y\right)+ \langle \mu^i | d {\bf c}_{i} \rangle \left(Y\right)\,,
\end{align*}
for every vector field $Y$. Here $d$ denotes the exterior differential in phase space, $\langle\, \cdot\, |\, \cdot\, \rangle$ is the usual pairing, and the $\lambda_{\mathrm{t}}^i$, $\lambda^i$,  $\mu_{\mathrm{t}}^i$, $\mu^i$ are Dirac multipliers. We get
\begin{align}\label{HVFHK}
 X^i_{\omega {\mathrm{t}}}   &=  \lambda^i_{\mathrm{t}}\,,&
 {\bf X}_{{\bf P} {\mathrm{t}} i}  \lcdr
    &=       \int_{\Sigma}     \cdot \, \mathrm{d}  \left( \epsilon_{ijk } e^j \wedge e^k \right)\,,\nonumber\\
 X^i_{\omega}           &= \lambda^i\,,         &
 {\bf X}_{{\bf P} i}  \lcdr 
     &=    2 \int_{\Sigma}  \cdot \wedge \mathrm{d} \left( \epsilon_{ijk }e^j_{\mathrm{t}} e^k\right)\,, \nonumber\\
 X^{i}_{e {\mathrm{t}}}   &=  \mu_{\mathrm{t}}^i\,,&
 {\bf X}_{{\bf p} {\mathrm{t}} i}   \lcdr &=  2  \int_{\Sigma} \epsilon_{ijk} \cdot e^j \wedge \mathrm{d} \omega^k\,,\nonumber\\
 X^{i}_{e}         &= \mu^i\,,      &
 {\bf X}_{{\bf p} i}  \lcdr          &= 2  \int_{\Sigma} \epsilon_{ijk} \cdot \wedge\left(  e^j \wedge \left(\lambda^k- \mathrm{d} \omega^k_{\mathrm{t}} \right) - e^j_{\mathrm{t}} \mathrm{d} \omega^{k} \right)\,.
\end{align}
The tangency condition of the Hamiltonian field \eqref{HVFHK} to the primary constrain submanifold defined by \eqref{consHKpb} gives:
\begin{subequations}\label{VectorField}
\begin{align}
0&=\bm{\imath}_X  { {d}} \left( {\bf C}_{{\mathrm{t}} i} \left( \cdot \right)\right)= {\bf X}_{{\bf P} {{\mathrm{t}} i}}   \left(\cdot\right)     \qquad  \Rightarrow  C^1_i:=  \mathrm{d} \left(\epsilon_{ijk } e^j \wedge e^k\right)  = 0\,, \\
0&=\bm{\imath}_X  { {d}}\left( {\bf C}_i \left( \cdot \right) \right)= {\bf X}_{{\bf P} i} \lcdr -2\int_{\Sigma} \epsilon_{ijk }  \cdot \wedge X^j_{e}\wedge e^k \nonumber\\
&  \qquad \qquad  \qquad \qquad \quad \qquad \qquad \Rightarrow  \mathrm{d} \left( \epsilon_{ijk }e^j_{\mathrm{t}} e^k\right) - \epsilon_{ijk}\mu^j \wedge e^k=0\,, \label{eqdm1}\\
0&=\bm{\imath}_X  { {d}} \left( {\bf c}_{{\mathrm{t}} i} \left( \cdot \right)\right)= {\bf X}_{{\bf p} {{\mathrm{t}} i}}   \left(\cdot\right) \,\qquad \,\,\, \Rightarrow   C^2_i:=\epsilon_{ijk} e^j \wedge \mathrm{d} \omega^k  = 0\,, \\
0&=\bm{\imath}_X  {{d}} \left( {\bf c}_i \left( \cdot \right) \right)= {\bf X}_{{\bf p} i} \lcdr \qquad \quad \,   \Rightarrow \epsilon_{ijk}\left[ e^j \wedge \left( \lambda^k- \mathrm{d} \omega^k_{\mathrm{t}} \right) - e^j_{\mathrm{t}} \mathrm{d} \omega^{k} \right]=0\,. \label{Diracm1}
\end{align}
\end{subequations}
Then, we have secondary constraints in the bulk, $C^1_i, C^2_i$, and equations for the Lagrange multipliers $\mu^i, \lambda^i$.  If we restrict ourselves to the non--degenerate case, using  Lemma \ref{lemma:Lms}, we find
\begin{subequations} \label{Dmhk}
\begin{align}
\mu^i&= \mathrm{d} e^i_{\mathrm{t}}  -\epsilon_{jk\ell}e_{\mathrm{t}}^k\left(\frac{e^\ell\wedge \mathrm{d} e^i}{\bm \omega}\right) e^j \,,\\
\lambda^i&=  \mathrm{d} \omega^i_{\mathrm{t}}   -\epsilon_{jk\ell}e_{\mathrm{t}}^k \left(\frac{e^\ell\wedge \mathrm{d} \omega^i}{\bm \omega}\right) e^j\,.
\end{align}
\end{subequations}
over the constraint surface. Next, we demand tangency to the submanifold defined by new constraints $C^1_i$, and $C^2_i$,
\begin{subequations}
\begin{align}
0&= \bm{\imath}_X  { {d}} \left( C^1_i\right) = \mathrm{d} \left( \epsilon_{ijk} \mu^j \wedge e^k\right) =  \mathrm{d} \left( \mathrm{d} \left( \epsilon_{ijk }e^j_{\mathrm{t}} e^k\right) \right)=0\,, \label{secohk1}\\
0&= \bm{\imath}_X  { {d}} \left( C^2_i\right) = \epsilon_{ijk}\left(  \mu^j \wedge \mathrm{d} \omega^k+ e^j\wedge \mathrm{d} \lambda^k \right) \,,  \label{secohk2}
\end{align}
\end{subequations}
where in the second equality of \eqref{secohk1} we have used \eqref{eqdm1}. Equation \eqref{secohk2} looks like an extra condition, but it is automatically satisfied; To see this, we calculate the differential of \eqref{Diracm1} obtaining $$ \epsilon_{ijk} e^j\wedge \mathrm{d} \lambda^k= \epsilon_{ijk}\left( \mathrm{d} e^i \wedge \left( \lambda^k- \mathrm{d} \omega^k_{\mathrm{t}} \right)- \mathrm{d} e^i_{\mathrm{t}} \wedge \mathrm{d} \omega^k \right).$$
Then, we substitute it in \eqref{secohk2}, and use the fact that the Dirac multipliers \eqref{Dmhk} can be written as  $\mu^i= \mathrm{d} e^i_{\mathrm{t}} +\imath_{\bm\varrho} \mathrm{d} e^i \,, \lambda^i=  \mathrm{d} \omega^i_{\mathrm{t}}   +\imath_{\bm\varrho} \mathrm{d} \omega^i$ with the vector field ${\bm \varrho}$ defined by $\imath_{\bm \varrho} e^i= e^i_{\mathrm{t}}$. This way we get
\begin{align*}
    \epsilon_{ijk}\left(  \mu^j \wedge \mathrm{d} \omega^k+ e^j\wedge \mathrm{d} \lambda^k \right)&=  \epsilon_{ijk}\left(  \imath_{\bm\varrho} \mathrm{d} e^i \wedge \mathrm{d} \omega^k+ \mathrm{d} e^i \wedge \imath_{\bm\varrho} \mathrm{d} \omega^k \right)  \\
    &=\imath_{\bm\varrho}  \left( \epsilon_{ijk}  \mathrm{d} e^i \wedge \mathrm{d} \omega^k \right)=\imath_{\bm\varrho} (0)=0\,.
\end{align*}
So, no new secondary constraints appear here. At the end, we have ${\bf C}_{\mathrm{t} i}(\cdot)$, ${\bf C}_{i}(\cdot)$, ${\bf c}_{ i}(\cdot)$, ${\bf c}_{\mathrm{t} i}(\cdot)$, $C^1_i$, $C^2_i$, which add up to $3+3+(3\times3)+(3\times3)+3+3=30$ constraints, and in the Hamiltonian vector field we have $ \lambda^i_{\mathrm{t}}$, $\mu^i_{\mathrm{t}}$, $\omega^i_{\mathrm{t}}$, $\omega^i$, $e^i_{\mathrm{t}}$, in total $3+3+3+3=12$ arbitrary functions. Therefore, the systems has $12$ first-class constraints, and $18$ second class constraints. On the other hand, we have, $\omega^i_{\mathrm{t}}$, $\omega^i$, $e^i_{\mathrm{t}}$, $e^i$, ${\bf P}_{\!\!\mathrm{t} i}(\cdot)$, ${\bf P}_{i}(\cdot)$, ${\bf p}_{\mathrm{t} i} (\cdot)$, ${\bf p}_{i}(\cdot)$, which are $(3+3\times3+3+3\times3)\times 2=48$ phase space variables. Then, the internal abelianized Husain-Kucha\v{r} model has $\frac{1}{2} \left(48-2\times 12-18\right)=3$  physical degrees of freedom (per point), the same as the $SO(3)$ Husain-Kucha\v{r} theory.

The equations for the integral curves of the Hamiltonian vector field \eqref{VectorField}, the part that contains the dynamics, for initial data on the constraint submanifold give
\begin{subequations}
\begin{align}
  \dot{\omega}^i & = \mathrm{d} \tau^i+\imath_{\bm\rho} \mathrm{d} \omega^i\,,\label{Adot}\\
  \dot{e}^i & = \mathrm{d} \rho^i+\imath_{\bm\rho} \mathrm{d} e^i\,,\label{edot}
\end{align}
\end{subequations}
where $\rho^i$ and $\tau^i$ are arbitrary functions of time (because the evolution of the $\omega_{\mathrm{t}}^i$ and $e_{\mathrm{t}}^i$ is arbitrary) and ${\bm \rho}$ is defined by $\imath_{\bm \rho} e^i=\rho^i$. In order to understand the meaning of \eqref{Adot} and \eqref{edot} one must to take into account that, by using Cartan's formula, the Lie derivative of $\omega^i$ and $e^i$ in the direction of ${\bm \rho}$ is
\begin{align*}
  &\pounds_{\bm \rho} \omega^i= \mathrm{d} \left( \imath_{{\bm \rho}} \omega^i \right) +\imath_{{\bm \rho}} \mathrm{d} \omega^i\,,\\
  &\pounds_{\bm \rho} e^i\,\,= \mathrm{d} \left( \imath_{{\bm \rho}} e^i \right) + \imath_{{\bm \rho}} \left( \mathrm{d} e^i\right)\,.
\end{align*}
Combining these expressions with \eqref{Adot} and \eqref{edot} we immediately get
\begin{subequations}
\begin{align}
  \dot{\omega}^i & =\pounds_{{\bm \rho}} \omega^i+ \mathrm{d} (\tau^i-\imath_{\bm \rho} \omega^i)\,,\label{Adot2}\\
 \dot{e}^i & =\pounds_{\bm \rho} e^i\,.\label{edot2}
\end{align}
\end{subequations}
The interpretation of the dynamics of the model is clear from \eqref{Adot2} and \eqref{edot2}:  for initial data satisfying the constraints, it is a combination of spatial diffeomorphisms and internal $U(1)^3$ rotations.

\section{A counterexample for the consistency of the Cartan-Palatini deformation}
\label{appendix_solution}

\noindent We want to find a solution to the equations of motion of the internally abelianized Cartan-Palatini action \eqref{HP_free}
\begin{align}
  \bd \be^I & =0\,, \label{eqs_e}\\
  \epsilon_{IJKL}\be^J\wedge \bd\bomega^{KL} & =0\,,  \label{eqs_omega}
\end{align}
for non-degenerate tetrads and satisfying 
\[
\bd(\be^J\wedge \bomega^I_{\phantom{I}J})\neq0\,.
\] 
We will assume that $\mathcal{M}$ can be covered with a single chart.
To begin with let us take $\be^I=\bd \bh^I$ with $\bh^I\in C^\infty(\mathcal{M})$ (for instance, $\bh^I=x^I$ where the $x^I$ are coordinates on $\mathcal{M}$) and $\bomega_{i0}=-\bomega_{0i}=0$, (with $i=1,2,3$).
By doing this we can write \eqref{eqs_omega} in the form
\begin{align*}
  \epsilon_{ijk}\bd \bh^i\wedge \bd\bomega^{jk} & =0\,, \\
  \bd \bh^0\wedge \bd\bomega^{ij} & =0\,.
\end{align*}
A non trivial solution to these equations is
\begin{align*}
\bomega^{ij}=\epsilon^{ij}_{\phantom{ij}k}\bh^0\bd \bh^k \ .
\end{align*}

Indeed, 
\begin{align*}
&\epsilon_{ijk}\bd \bh^i\wedge \bd \bomega^{jk}=\epsilon_{ijk}\bd \bh^i\wedge\epsilon^{jk}_{\phantom{jk}l}\bd \bh^0\wedge \bd \bh^l=2\bd \bh^i\wedge \bd \bh^0\wedge \bd \bh_i=0\,,\\
&\bd \bh^0\wedge \bd \omega^{ij}=\bd \bh^0\wedge\epsilon^{ij}_{\phantom{ij}k}\bd \bh^0\wedge\bd \bh^k=0\,.
\end{align*}
On the other hand
\[
\bd(\be^J\wedge \bomega^i_{\phantom{i}J})=\bd (\be^j\wedge \bomega^i_{\phantom{i}j})=\bd (\bd \bh^j\wedge\epsilon^i_{\phantom{i}jk}\bh^0\bd \bh^k)=-\epsilon^i_{\phantom{i}jk}\bd \bh^j\wedge\bd \bh^0\wedge \bd \bh^k\neq0\,.
\]

\section{Linearization and abelianization of the Cartan-Palatini action in 3+1 dimensions}\label{lacp}

Let $\bh^{I}$ four smooth scalar fields chosen in such a way that the $\bd \bh^I$ define a non- degenerate coframe (notice that $g=\eta_{IJ}\bd \bh^{I}\otimes \bd \bh^{J}$ is the Minkowski metric). Let us introduce the expansions 

\begin{subequations}
\begin{align}
\be^{I}&=\bd \bh^{I}+\varepsilon \Tilde{\be}^{I}+\varepsilon^2 \Tilde{\be}^{I}_2+\cdots  \,,\\
\bomega^{I}{}_J&=0+\varepsilon \Tilde{\bomega}^{I}{}_J+\varepsilon^2 \Tilde{\bomega}_{2\,\,\,J}^{\,\,\,I}+\cdots \,,  
\end{align}
\end{subequations}
where $\varepsilon$ is a (small) real parameter. To second order in $\varepsilon$ the Cartan-Palatini Lagrangian becomes
\begin{align*}
\epsilon_{I J K L} \be^{I} \wedge \be^{J} \wedge \bF^{K L}=&\varepsilon^{2} \epsilon_{IJKL}\left( 2 \bd \bh^{I} \wedge \Tilde{\be}^{J} \wedge \bd \Tilde{\bomega}^{K L}+ \bd \bh^{I} \wedge \bd \bh^{J} \wedge \Tilde{\bomega}^{K}{}_P \wedge \Tilde{\bomega}^{P L}\right)+O(\varepsilon^3)\,,\nonumber
\end{align*}
where we have dropped all the total derivatives.

The Abelian action can be taken as
\begin{equation}
    S^0(\Tilde{\be}, \Tilde{\bomega})= \int_{\mathcal{M}} \epsilon_{IJKL} \bd \bh^{I} \wedge \Tilde{\be}^{J} \wedge \bd \Tilde{\bomega}^{K L}\,. \label{new5}
\end{equation}
Notice that integrating by parts and making the change of variables $\epsilon_{IJKL}\bd \bh^J\wedge \Tilde{\bomega}^{KL}=:\bB_I$, the action \eqref{new5} can be written as a BF theory (for a similar idea see \cite{BarrosESa}),  \[S^0(\Tilde{\be}, \bB_I)= \int_{\mathcal{M}} \bB_I \wedge \bd \Tilde{\be}^I\,,\] 
which shows it topological character. This can also be explicitly seen by studying the field equations corresponding to \eqref{new5}
\begin{subequations}
\begin{align}
 \epsilon_{IJKL} \bd \bh^{I} \wedge \bd \Tilde{\be}^{J}  &=0 \,, \label{new1}\\
  \epsilon_{IJKL} \bd \bh^{J} \wedge \bd \Tilde{\bomega}^{K L}  &=0  \,. \label{new2}
\end{align}
\end{subequations}
Using that the $\bd \bh^{I}$ are non-degenerate, equation \eqref{new1} implies
\begin{equation}\label{solE3A}
   \bd \Tilde{\be}^{J}=0 \Rightarrow  \Tilde{\be}^{J}= {\bd}{\bm{\mathrm{f}}}^I+\sum_{i=1}^{n_1}{\bm{\mathrm{\eta}}}_i^I \phi^{\scriptscriptstyle (1)}_i\,,
\end{equation}
for some functions $\bm{\mathrm{f}}^I$, ${\bm{\mathrm{\eta}}}_i^I\in \mathbb{R}$, and the representatives of the de Rham cohomology classes $\phi^{\scriptscriptstyle (1)}_i$ (which are closed 1-forms).

From \eqref{new2} we get 
\begin{align}
    \bd \left(   \epsilon_{IJKL} \bd \bh^{J} \wedge \Tilde{\bomega}^{K L} \right)=0
\end{align}
hence, we have
\begin{equation}
\epsilon_{IJKL}\bd \bh^J\wedge \Tilde{\bomega}^{KL}={\bd}{\bm{g}}_I+\sum_{j=1}^{n_2}{\bm{\gamma}}_{Ij}\phi^{\scriptscriptstyle (2)}_j=:\bG_I\,, \label{new3}
\end{equation}
where  ${\bm{g}}_I\in\Omega^1(\mathcal{M})$, ${\bm{\gamma}}_{Ij}\in\mathbb{R}$ and the closed 2-forms $\phi^{\scriptscriptstyle (2)}_j\in\Omega^2(\mathcal{M})$ are representatives of the de Rham cohomology classes in $H^2_{\mathrm{dR}}(\mathcal{M})$. To solve for $\Tilde{\bomega}^{KL}$  in \eqref{new3}, we use Lemma \ref{lemma:index_eq} with $\mathsf{vol}_{\bd \bh}$. We thus obtain
\begin{align}\label{solE7}
    \Tilde{\bomega}^{KL} = -\frac{1}{2} \left( \frac{\bG_I\wedge \bd \bh^K \wedge \bd \bh^L}{\mathsf{vol}_{\bd \bh}}\right)  \bd \bh^I -\frac{1}{2}  \left(\frac{\bG_I\wedge \bd \bh^I \wedge \bd \bh^{[K}}{\mathsf{vol}_{\bd \bh}} \right) \bd \bh^{L]}  \ .
\end{align}
Plugging \eqref{solE3A} and \eqref{solE7} into the symplectic form one can see that the action \eqref{new5} only describes topological degrees of freedom.

\bibliographystyle{JHEP}
\bibliography{main} 

\end{document}